\documentclass[reprint,amsmath,amssymb,aps,prd,groupedaddress,nofootinbib,twocolumn,superscriptaddress]{revtex4-1}
\usepackage{graphicx}
\usepackage{amsthm,amssymb,amsmath,braket,mathdots}
\usepackage{bm}
\usepackage[pagebackref=false,pdfnewwindow=true]{hyperref} %\hypersetup{draft}
\usepackage{epstopdf,psfrag}
\usepackage{relsize,amsbsy}
\usepackage[export]{adjustbox}
\usepackage{float}
\usepackage{subfigure}
\usepackage{color}
\usepackage{wasysym}

\usepackage{graphicx,xcolor,tikz}

\DeclareMathOperator{\sgn}{sgn}
\DeclareMathOperator{\tr}{Tr}
\DeclareMathOperator{\re}{Re}
\DeclareMathOperator{\im}{Im}

\newcommand{\diff}{\mathrm{d}}

\newcommand{\be}{\begin{equation}}
\newcommand{\ee}{\end{equation}}

\newcommand{\bit}{\begin{enumerate}}
\newcommand{\eit}{\end{enumerate}}

\newcommand{\non}{\nonumber}
\newcommand{\dg}{\dagger}
\newcommand{\ve}[1]{{\bf #1}}
\newcommand{\e}{\epsilon}
\newcommand{\df}[1]{|\vec{\mathcal{F}}_{#1}|}
\newcommand{\nf}[1]{\vec{\mathcal{F}}_{#1}}
\newcommand{\kv}{\ve{k}}
\newcommand{\qv}{\ve{q}}

\definecolor{bananayellow}{rgb}{1.0, 0.88, 0.21}
\definecolor{straw}{rgb}{0.32, 0.28, 0.1}

\allowdisplaybreaks
\begin{document}
\title{ Phonon dynamics in the Kitaev spin liquid }
\author{Mengxing Ye}
\affiliation{Kavli Institute for Theoretical Physics, University of
California, Santa Barbara, CA 93106, USA}
\affiliation{School of Physics and Astronomy, University of Minnesota, Minneapolis, MN 55455, USA}
\author{Rafael M. Fernandes}
\affiliation{School of Physics and Astronomy, University of Minnesota, Minneapolis, MN 55455, USA}
\author{Natalia B. Perkins}
\affiliation{School of Physics and Astronomy, University of Minnesota, Minneapolis, MN 55455, USA}
\date{\today} 
		
\begin{abstract}
The search for fractionalization in quantum spin liquids largely relies on their decoupling with the environment. However, the spin-lattice interaction is inevitable in a real setting. While the Majorana fermion evades a strong decay due to the gradient form of spin-lattice coupling, the study of the phonon dynamics may serve as an indirect probe of  fractionalization of  spin degrees of freedom. Here we propose that the signatures of fractionalization can be seen in the sound attenuation and the Hall viscosity. Despite the fact that both quantities can be related to the imaginary part of the phonon self-energy, their origins are quite different, and the time-reversal symmetry breaking is required for the Hall viscosity. First, we compute the sound attenuation due to a phonon decay by scattering with a pair of Majorana fermions and show that it is linear in temperature ($\sim T$). We argue that it has a particular angular dependence providing the information about the spin-lattice coupling and the low-energy Majorana fermion spectrum. The observable effects in the absence of time-reversal symmetry are then analyzed. We obtain the phonon Hall viscosity term from the microscopic Hamiltonian with time-reversal symmetry breaking term. Importantly, the Hall viscosity term mixes the longitudinal and transverse phonon modes and renormalize the spectrum in a unique way, which may be probed in spectroscopy measurement.  

%We propose that the phonon dynamics in the Kitaev spin liquid can provide an important information about the fractionalization of the spin degrees of freedom. 
\end{abstract}
	
\maketitle		

\section{Introduction} 
% Add a discussion about the spin-lattice coupling is inevitable but can be important.

Quantum spin liquids (QSLs), a particularly fascinating class of frustrated magnets,  have been a focus of  condensed matter research
since the initial proposal ~\cite{Anderson1973}. These systems evade magnetic order down to zero temperature and harbor a remarkable set of collective phenomena, including topological ground-state degeneracy, long-range entanglement, and fractionalized excitations \cite{XGWen2002,Kitaev2006,Balents2010,Savary2016}. Of particular interest is the Kitaev honeycomb model which describes a system of spin-1/2 at sites of a honeycomb lattice interacting via Ising-like frustrated nearest-neighbor exchange interactions \cite{Kitaev2006}. This model is not only exactly solvable with a QSL ground state, but is also realizable in real materials~\cite{Jackeli2009,Trebst2017,Winter2016,Winter2017,hermanns2018physics,Takagi2019,Motome2019}.

\begin{figure*}[t]
\centering
\subfigure[]
{\includegraphics[width=0.55\columnwidth]{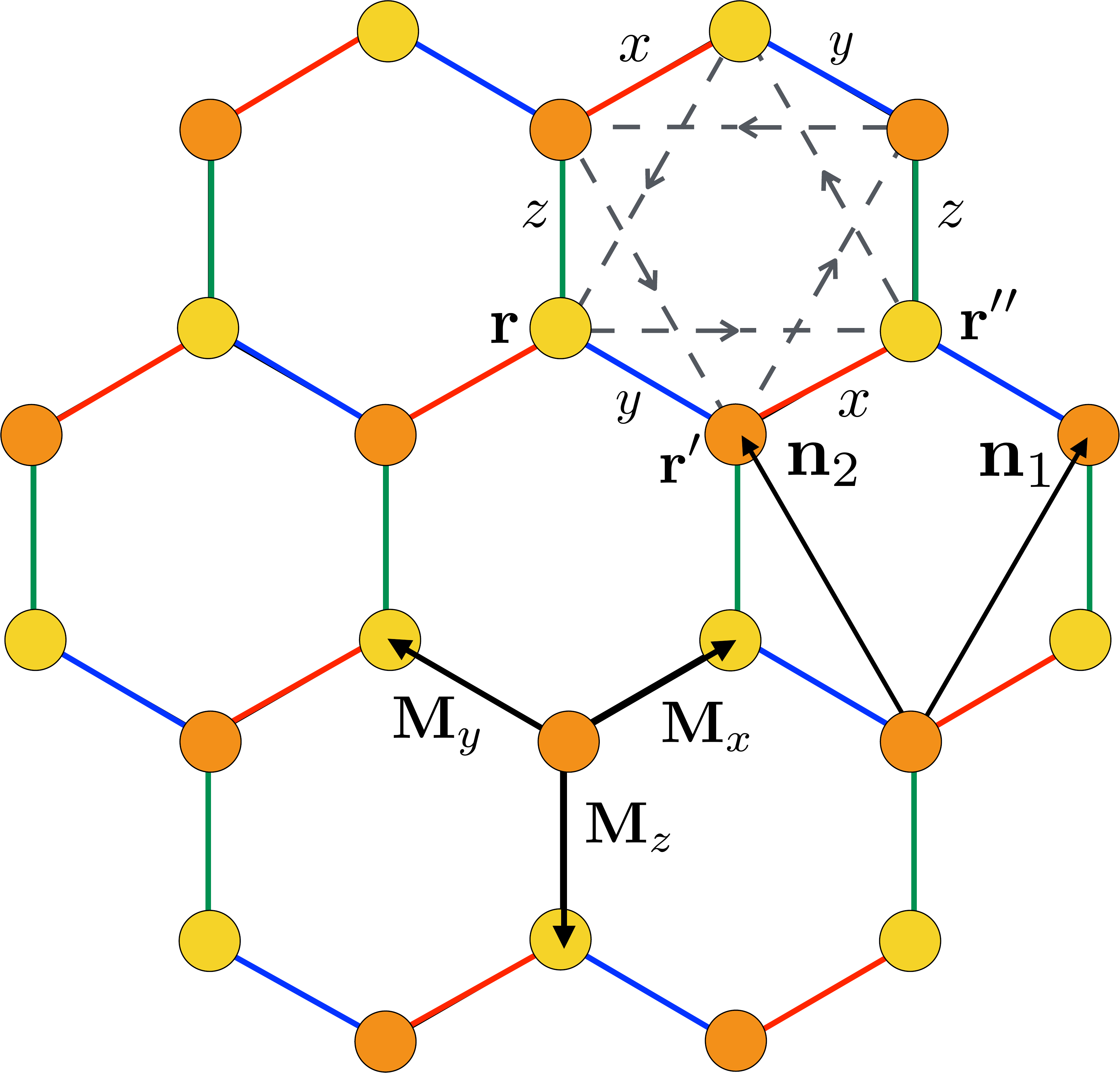}}\quad
\subfigure[]
{\includegraphics[width=0.5\columnwidth]{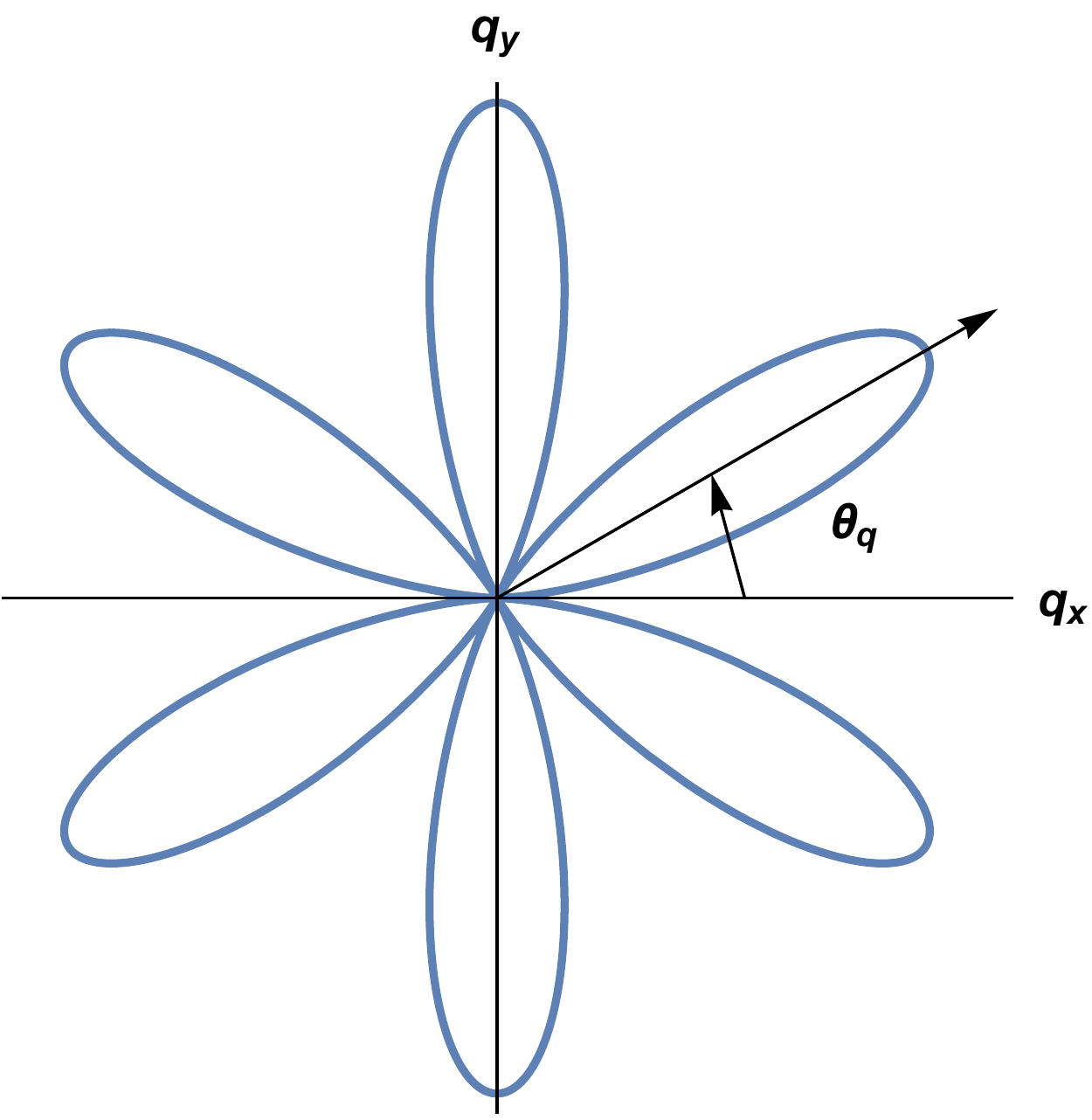}}\quad
\subfigure[]
{\includegraphics[width=0.85\columnwidth]{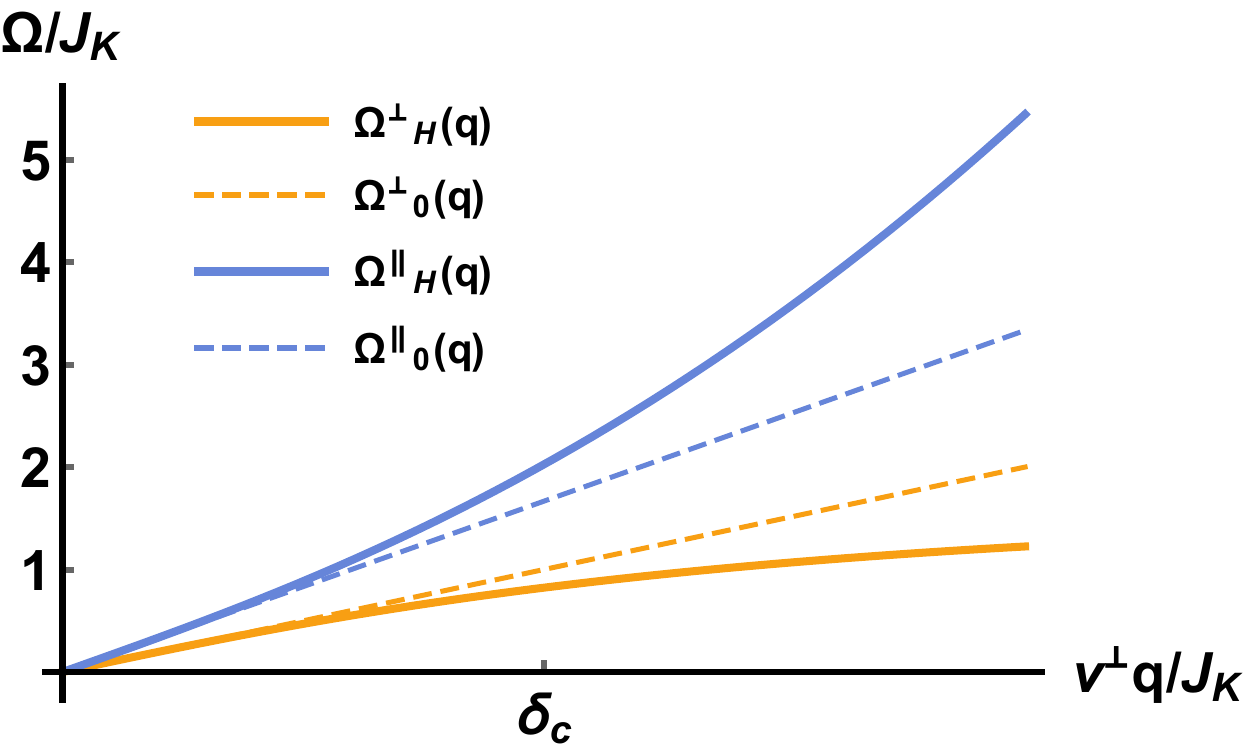}}
\caption{(a) Kitaev model on the honeycomb lattice. The sites in
the two sublattices of the honeycomb lattice  ($A$ and $B$) are marked by yellow  and orange circles.   Here ${\bf n}_1=(\frac{\sqrt{3}}{2},\frac{3}{2})$ and ${\bf n}_2=(-\frac{\sqrt{3}}{2},\frac{3}{2})$  are two unit vectors, and we set the lattice constant $\ell_a=1$. The three vectors $\mathbf M_{x,y} = (\pm\frac{\sqrt{3}}{2},\frac{1}{2})$, and $\mathbf M_z = (0, -1)$ connect nearest neighbors by $x$, $y$, and $z$ bonds, respectively. The three sites ${\bf r},\,{\bf r}', and {\bf r}''$ form a representative three-spin link $\langle{\bf r},\,{\bf r}',\,{\bf r}''\rangle_{yx}$ as defined below Eq.~\eqref{eq:Kmodel1}.
 (b) Angular dependence of the sound attenuation coefficient $\alpha^\parallel_s$ for the longitudinal mode.
 %: radial variable for the magnitude of $\alpha_s$, angular variable for the angle of phonon momentum $\bf{q}$.
$\alpha^\perp_s$ for the transverse mode follows the same shape but is rotated by $\pi/6$. (c) Acoustic phonon spectrum $\Omega(\ve{q})$. The dashed lines depict the bare  spectrum with longitudinal and transverse modes. Solid lines depict the spectrum renormalized due to the mixing of transverse and longitudinal modes when time-reversal symmetry is broken. Here $\delta_c$ denotes the characteristic dimensionless scale beyond which the phonon spectrum is significantly bent [see Eq.~\eqref{eq:mixingc}].
}
\label{fig:summary}
\end{figure*}

Unfortunately, the ground states of these systems  cannot be easily captured by experiment, remaining featureless to conventional local  probes. A promising route to detect QSLs is to look for signatures of fractionalization in dynamical probes, such as inelastic neutron scattering (INS)~\cite{Nagler91,Tennant93,Knolle2015,Knolle2014a,Banerjee2016,Banerjee2017},  Raman scattering~\cite{Ko2010,Sandilands2015,Knolle2014,Nasu2016,Rousochatzakis2019,Sahasrabudhe2019,Wulferding2019}, resonant inelastic x-ray scattering (RIXS)~\cite{Gabor2016,Halasz2019}, ultrafast spectroscopy~\cite{Alpichshev15}  and two-dimensional (2D) terahertz non-linear coherent spectroscopy~\cite{Wan2019}.

Here we propose to study the signatures of the fractionalized excitations in Kitaev materials   by exploiting  their coupling to lattice vibrations. Indeed, it is well known that the coupling between collective spin and lattice degrees of freedom plays an important role in the fundamental properties of correlated materials, and in 
many cases, it has been explored rather carefully.  For example, attenuation of sound due to electron-phonon coupling has been widely used to study the electronic properties  and phase transitions of various solids with complex order parameters~\cite{Pippard1955,Akhiezer1957,Blount1959} and was proved to provide useful information on both  normal and superconducting states~\cite{Tsuneto1961,Ott1985,Kazumi1994}. The magneto-elastic coupling also plays an important role in probing vestigial phases of frustrated magnets \cite{Fernandes19}. 
The possibility to use the sound attenuation to study the nature of two-dimensional frustrated magnetic systems, and in particular QSLs, has also been recently discussed in the literature~\cite{Kreisel2011,Plee2011,Plee2013,Brenig2019}. In particular, the spin-lattice coupling was shown to play an essential role in the theoretical interpretation of some experiments in the study of Kitaev materials, e.g.\ the thermal Hall transport measurement~\cite{Kasahara2018,Ye2018b,Rosch2018}. This suggests that, in a realistic setting, the spin-lattice coupling in Kitaev materials could be large enough to affect the phonon dynamics \cite{Brenig2019}. However, a detailed analysis of the sound attenuation in Kitaev magnets, and in particular the signatures constrained by the symmetry of the spin-lattice coupled system, remain mostly unexplored. 
 
In this paper we argue that the phonon dynamics   can be used to probe spin fractionalization in Kitaev materials, and  in particular in $\alpha$-RuCl$_3$~\cite{Plumb2014,Banerjee2017,Do2017,Hirobe2017,Kasahara2018}.  
 As a proof of principle, we study the pure Kitaev model~\cite{Kitaev2006} in the isotropic limit on a 2D honeycomb lattice [see Fig.~\ref{fig:summary} (a)], with parameters such as sound velocity $v_s$ and Kitaev interaction $J_K$ extracted from experimental data on $\alpha$-RuCl$_3$~\cite{Winter2016,Hirobe2017}.
% J_K = -20 meV
% v_s = 1490 m/s
In the pure Kitaev model, the spins fractionalize
into two types of elementary excitations --
Majorana fermions and emergent gauge fluxes~\cite{Kitaev2006}.  The ground state  of the isotropic Kitaev  spin liquid corresponds to  a 
fixed zero-flux configuration. Therefore at  temperatures below the flux gap $\Delta_{\mathrm{flux}}$, the low-energy magnetic excitations  are solely dispersive Majorana fermions.
We assume that the Majorana fermion-phonon coupling arises from the fact that the
 Kitaev interaction $J_K$ depends on the  relative positions between the spins. 
 In order to obtain an effective low-energy theory, we perform a microscopic analysis of  the change of  the spin exchange energy due to the lattice distortion and obtain the explicit form of the Majorana fermion-phonon coupling  by considering acoustic phonon modes coupled
to low-energy spin degrees of freedom expressed in terms of Majorana fermions.
%  We show that the Majorana fermion-phonon  coupling in the Kitaev materials leads to an acoustic phonon decay.
We also find that in the low-energy limit, this  coupling has essentially the same  form as  that  obtained  from the symmetry considerations in Ref.~\cite{Plee2013} for algebraic spin liquids.

In this picture, we study the observable consequences of the spin-lattice coupling through the phonon dynamics. By calculating the phonon self-energy,  we compute the sound attenuation coefficient and the Hall viscosity, analyzing their observable signatures.
%, the renormalization to the sound velocity and describe the mixing of the transverse and longitudinal phonon modes.

\begin{figure}[t]
\includegraphics[width=1\columnwidth]{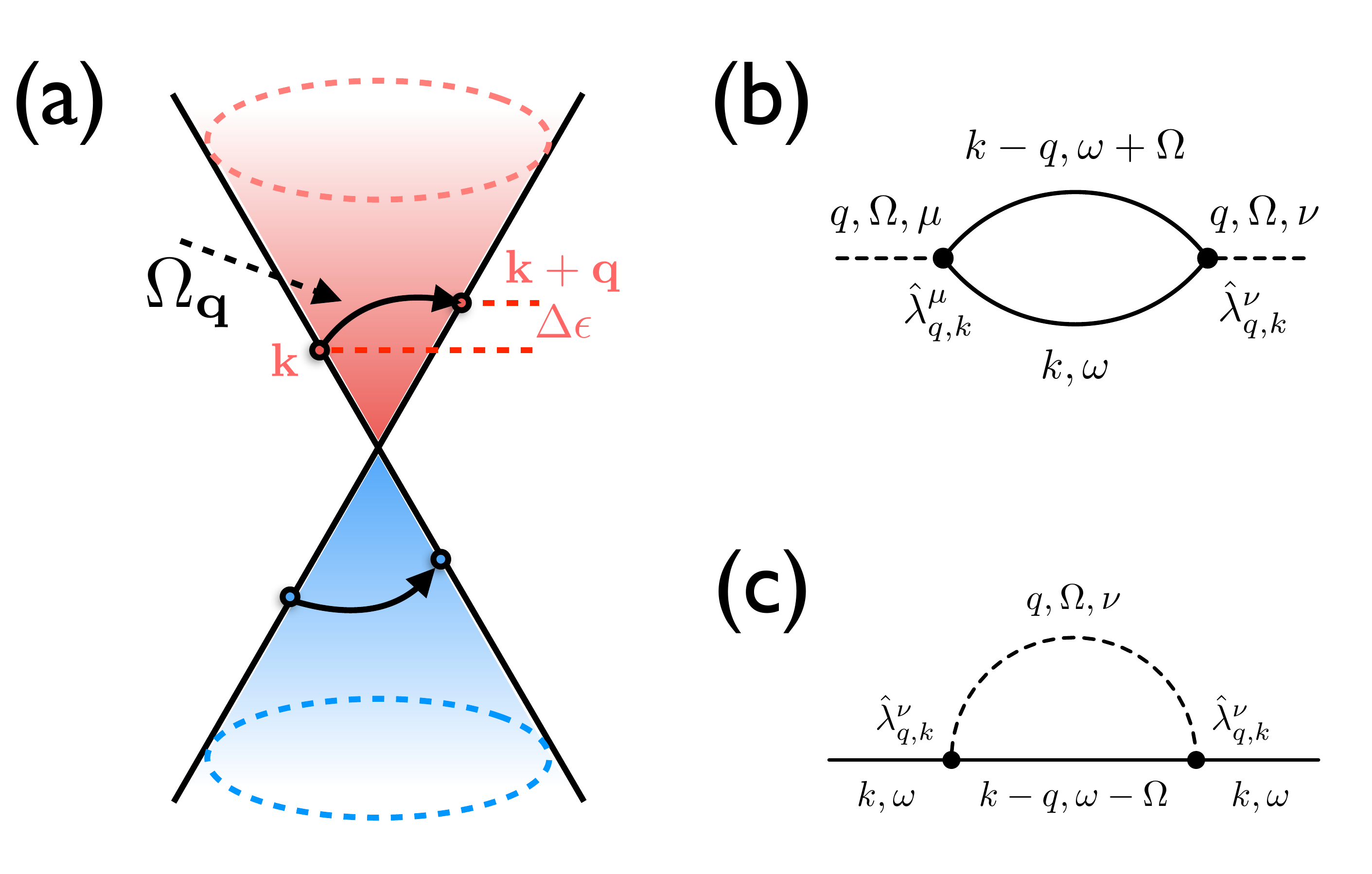}
%\begin{minipage}[c][4.5cm][t]{.35\columnwidth}
%  \vspace*{\fill}
%  %\centering
%	\includegraphics[width=1\columnwidth]{Fig2a}
%	\vspace*{\fill}
%\end{minipage}%
%\begin{minipage}[c][4.5cm][t]{.65\columnwidth}
% % \vspace*{\fill}
%  \centering
%	\includegraphics[width=0.9\columnwidth]{Bubble.png}
%	\includegraphics[width=0.9\columnwidth]{OneLoop.png}
%	%\vspace*{\fill}
%\end{minipage}
\caption{(a) On-shell process in which a phonon mode with $\Omega_{\bf q}=v_s |{\bf q}|$ (black dash arrow) excites a Majorana fermion from ${\bf k}$ to ${\bf k+q}$ such that $\Delta\epsilon=\epsilon_{\bf k+q}-\epsilon_{\bf k}=\Omega_{\bf q}$.
 Red (blue) shades indicate thermal excitation (depletion) probability of Majorana fermions near a Dirac cone at $\pm K$.  (b) Feynman diagram representing  the phonon self-energy Eq.~\eqref{Appeq:Polarization}, which determines the sound attenuation and Hall viscosity for a given set of phonon polarization indices $\mu,\nu$.
  (c) Feynman diagram representing  the Majorana fermion self-energy Eq.~\eqref{Appeq:MFSE} used to find the Majorana fermion lifetime due to spin-lattice coupling.} % \, \centering; da yuan
\label{fig:Attenuation}
\end{figure}

We first show that the sound attenuation is determined by the decay of a phonon due to scattering with a pair of Majorana fermions
[see Fig.~\ref{fig:Attenuation}(a)], with the attenuation rate linear in temperature due to the vanishing density of states at the Dirac points. Importantly, this  is the dominant process compared with the sound attenuation due to, e.g., phonon-phonon interactions that scale as proportional to $ T^3$~\cite{Woodruff1961}. 
Moreover, we find that due to the anisotropic form of  the Majorana fermion-phonon  coupling
and the Dirac fermion like low-energy Hamiltonian of the magnetic excitations, 
the sound attenuation  shows  a strong angular dependence at the leading order in  phonon momentum $q$ [see Fig.~\ref{fig:summary}(b)]. Consequently, the angular dependence may offer a quite powerful probe of  Majorana fermions with Dirac spectrum in Kitaev materials. 

The same Majorana fermion-phonon interaction  also gives rise to  the finite life-time of the Majorana fermions. This effect, however,  is quite weak, and the life-time scales as $\tau_{f}\sim\im \Sigma_{MF}\sim T^2$, which is much smaller than the typical fermion energy $\sim T$.

We next study the modifications of the phonon dynamics in the absence of time-reversal symmetry due to applying a magnetic field $h$. Assuming the phonons do not couple to the magnetic field directly, these corrections are induced by the spin-lattice coupling. While the sound attenuation coefficient $\alpha_s$ is not changed qualitatively in a small magnetic field when $(h/J_K)^3\lesssim q \ell_a$, where $\ell_a$ is the lattice constant, the Majorana fermion-phonon coupling introduces another interesting effect to the phonon system, i.e.\ a Berry phase term that mixes the transverse and longitudinal phonon modes.
It is encoded in the phonon long-wavelength effective action as the Hall viscosity term, 
%which is non-dissipative~\cite{Avron1995,Barkeshli2012,Rosch2018}.
%In terms of {\cred the phonon low-energy} effective action, 
which
is the leading order term breaking the time-reversal symmetry
%, i.e.\ the Hall viscosity term
~\cite{Avron1995,Barkeshli2012}. It is associated with the non-dissipative response to a velocity gradient of a fluid~\cite{Avron1995,Read2011,Barkeshli2012}, and also contributes to the thermal Hall effect~\cite{Shi2012,Rosch2018}.
As it comes from off-shell processes, both the high-energy and low-energy Majorana fermion modes contribute. In our analysis presented below, we calculate the Hall coefficient from the whole Majorana fermion spectrum. We compute the non-perturbative contribution in $h$ to the Hall viscosity coefficient, which matches the result in an infinitesimally small field. We also compute the perturbative correction and show that the Hall coefficient decreases as the magnetic field strength increases.
We also note that the mixing renormalizes the phonon spectrum prominently above a characteristic phonon momentum $q_c$ [see Fig.~\ref{fig:summary}(c)] that may be observed in some spectroscopy measurements.

The rest of the paper is organized as follows.
In Sec.~\ref{Sec:model}, we  present the spin-phonon Hamiltonian. First, we discuss various aspects of the extended Kitaev model (including the three-spin interaction $\kappa$-term that breaks time-reversal symmetry) 
that are most relevant for this study. Second, we  introduce the lattice Hamiltonian for the acoustic phonons on the honeycomb lattice. Third, perform the symmetry analysis of the magneto-elastic coupling. The explicit microscopic derivation of the coupling vertices in terms of Majorana fermions is later presented in Sec.~\ref{Sec:MFPhcoupling}.
%
%The explicit mi- croscopic derivation of the coupling vertices using Majo- rana fermion representation of spin is later presented in Sec. IV.
 % 
In Sec.~\ref{Eltheory}, we describe the dynamics of acoustic phonons both in the presence and in the absence of the time reversal symmetry using the long-wavelength effective action approach.
To obtain the hydrodynamic coefficients from the microscopic Hamiltonian, we use the diagrammatic techniques and compute the phonon polarization bubble. Details of the calculation of the bare Green's function and self-energy are presented in Sec.~\ref{SMphonons}.
%integrate out the Majorana fermions and obtain an effective low-energy theory only for the phonons. For calculations of the phonon polarization bubble at finite temperature  we use the Matsubara frequency representation.
%
In Sec.~\ref{sec:Attenuation}, 
we relate the imaginary part of the diagonal components of the phonon polarization bubble to the attenuation coefficients. 
%we present the analytical calculation of attenuation coefficient and relate it to the imaginary part of the diagonal components of the phonon polarization bubble.
%
The phonon dynamics in the system  with time-reversal-symmetry breaking is discussed in Sec.~\ref{Sec:phonTRSbr}.
We first relate the off-diagonal component of the phonon polarization bubble to the Hall viscosity coefficient, and then show how it renormalizes the phonon spectrum. Both the perturbative and non-perturbative corrections in terms of $\kappa$ are obtained. 
%Here, we first derive the Hall viscosity coefficient and relate it to the off-diagonal component of the phonon polarization bubble, and then show how it renormalizes the phonon spectrum. Both the perturbative and non-perturbative corrections in terms of $\kappa$ are obtained. 
%
A summary and a general discussion are given in Sec.~\ref{Sec:summary}. 
Auxiliary information and technical details are provided in the Appendixes.
%~ \ref{MajoranaGF},~\ref{Appsec:alphas} and \ref{Appsec:Attenuation}.

%%%%%%%%%%%%%%%%%%%%%%%%%%%%%%%%%%%%%%%%%%%%%%%%%%%%%%%%%
\section{The model} \label{Sec:model}
We focus our discussion on the spin-phonon Hamiltonian
\begin{align}
\label{eq:model}
H=H_{\text s}+H_{\mathrm{ph}}+H_{\text c}.
\end{align} 
 
The {\it first term} in Eq. (\ref{eq:model}) is  the spin  Hamiltonian given by
\begin{align}
H_{\text s}  = -\sum_{\alpha,{\bf r}\in A}  J^\alpha  \sigma_{\bf r}^{\alpha} \sigma_{{\bf r}+{\bf M}_\alpha}^{\alpha} - 
\kappa \sum_{\langle {\bf r},{\bf r}',{\bf r}'' \rangle_{\alpha\gamma}}\sigma^\alpha_{\bf r} \sigma^\beta_{{\bf r}'} \sigma^\gamma_{{\bf r}''},
\label{eq:Kmodel1}
\end{align}
where $J^{\alpha}$  denotes the nearest neighbor Kitaev interaction  on  the corresponding bond of type $\alpha=x,y,z$,  $\sigma^\alpha_{\bf r}$ are the Pauli matrices, and ${\bf M}_\alpha$ labels the three inequivalent bonds on the honeycomb lattice [see Fig.~\ref{fig:summary} (a)]. In the following, we ignore  the presence of weak non-Kitaev couplings and  weak anisotropy of the Kitaev interactions in real materials, and assume  $J^\alpha=J_K$  and  the $C_{6v}$ symmetry of the honeycomb lattice. The $\kappa$ term breaks time-reversal and mirror ($\sigma_v$) symmetry while preserving the exact solubility of the model. The three-spin link notation $\langle {\bf r},{\bf r}',{\bf r}'' \rangle_{\alpha\gamma}$ labels bonds $\ve{r}\ve{r'}$ and $\ve{r'}\ve{r''}$ of types $\alpha$ and $\gamma$, respectively, on three adjacent sites ${\bf r},{\bf r}'$ and ${\bf r}''$ moving counterclockwise [see Fig.~\ref{fig:summary} (a)]. Here $\alpha,\beta$, and $\gamma$ are determined such that $\beta\neq \alpha,\gamma$. It may be considered as the leading order perturbative effect of the magnetic field in the zero flux sector~\cite{Kitaev2006}, and $\kappa\sim \frac{h_x h_y h_z}{J_K^2}$.
%; \textcolor{blue}{it appears when an external magnetic field with out-of-plane component is applied}

Decomposing spin operators into two Majorana fermions $\sigma_{\bf r}^{\alpha}=i c_{\bf r} b_{\bf r}^{\alpha}$~\cite{Kitaev2006}, 
the spin Hamiltonian can be written as 
\begin{align}
\label{eq:Kmodels}
H_{\text s}\rightarrow \tilde H_{\text s}=&J_K\sum_{\alpha,\ve r\in A}iu^{\alpha}_{\ve r,\ve r+\ve{M}_{\alpha}}c_{\ve r}c_{\ve r+\ve{M}_{\alpha}} \\\nonumber &+\kappa\sum_{\langle {\bf r},{\bf r}',{\bf r}'' \rangle_{\alpha\gamma}} i u^{\alpha}_{\ve r,\ve r'} u^{\gamma}_{\ve r',{\bf r}''}  c_{\ve r} c_{{\bf r}''},
\end{align}
where $u^{\alpha}_{\ve r,\ve r+\ve{M}_{\alpha}}=i b^{\alpha}_{\ve r}b^{\alpha}_{\ve{r}+\ve{M}_\alpha}$, ${\bf r}'=\ve r+\ve{M}_{\alpha}$, and ${\bf r}''=\ve r+\ve{M}_{\alpha}-\ve{M}_{\gamma}$. Here 
 we use $\tilde H_{\text s}$ to label the Hamiltonian in the enlarged Hilbert space in terms of Majorana fermions. Importantly, the gauge dependent link variables $u^{\alpha}_{\ve r,\ve r+\ve{M}_{\alpha}}$ are mutually commuting constants of motion in $\tilde H_s$, and they give rise to physical static $Z_2$-flux degrees of freedom $W_p=\Pi_{\langle\ve{r},\ve{r}+\ve{M}_\alpha\rangle\in p}u^{\alpha}_{\ve{r},\ve{r}+\ve{M}_\alpha}=\pm 1$ at the plaquette $p$. Each flux sector can be characterized by a static configuration of $\{u^{\alpha}_{\ve r,\ve r+\ve{M}_{\alpha}}=\pm 1\}$,  for which one obtains the free-fermion Hamiltonian in terms of $c_{\ve r}$. 
Thus, the Majorana fermion Hamiltonian in the ground state zero-flux sector can be obtained by setting all $u^{\alpha}_{\ve r,\ve r+\ve{M}_{\alpha}}=1$. In momentum space, it reads
\begin{align}  
\tilde H_{\text s}=\frac{1}{2}\sum_{\bf k} 
\psi^T_{-\ve{k}}
(-\tau_x\im f_{\bf k}-\tau_y\re f_{\bf k} +\tau_z \Delta_{\bf k})
\psi_{\ve k},
\label{eq:Kmodelf}
\end{align}
where $\psi_{\ve{k}}=(\frac{1}{\sqrt{2N}}\sum_{{\ve r}\in A}e^{i\ve{k}\cdot\ve{r}}c_{\ve{r},A},\frac{1}{\sqrt{2N}}\sum_{{\ve r}\in B}e^{i\ve{k}\cdot\ve{r}}c_{\ve{r},B})^T$  are the complex fermions obtained from the Fourier transform of  the Majorana fermions $c_{A,\ve{r}}$ and $c_{B,\ve{r}}$
on sublattices A and B, respectively, and $N$ is the number of unit cells, 
$\tau_{x,y}$ are the auxiliary Pauli matrices in sublattice space, $f_\kv=2J_K(1+2\cos \sqrt{3}k_x/2\, e^{i3k_y/2})$ from the nearest neighbor Majorana fermion hopping, and $\Delta_{\bf k}=4\kappa \big(\sin \ve k\cdot\ve n_1-\sin \ve k\cdot\ve n_2+\sin \ve k\cdot (\ve n_2-\ve n_1)\big)$ comes from the second neighbor Majorana hopping due to the time-reversal breaking part of the spin Hamiltonian, with ${\bf n}_1=(\frac{\sqrt{3}}{2},\frac{3}{2})$ and ${\bf n}_2=(-\frac{\sqrt{3}}{2},\frac{3}{2})$.
%\MY{[MY: In the numerical computation, I actually used $\Delta_{\bf k}=\kappa \big(\sin \ve k\cdot\ve n_1-\sin \ve k\cdot\ve n_2+\sin \ve k\cdot (\ve n_2-\ve n_1)\big)$, i.e. there is a mistake of factor of 4. The factor doesn't change result of Eq. 53 because it only depends on the sign of $\kappa$, but the expression of $r(\kappa)$ needs to be rescaled. Please see the new Fig.~\ref{fig:rk} for the updates. It basically compress the longitudinal ($\kappa/J$) axis by a factor of 4.]}
When $\kappa=0$, the Majorana fermion spectrum $\epsilon_\kv\sim |f_\kv|$ contains two Dirac points at the corners of the Brillouin zone $\pm K$. 
% Note that though we discuss the solution for Eq.~\eqref{eq:Kmodel1}, this exact parton construction with \emph{static} $Z_2$ flux works for a generic Kitaev coupling configuration, even when translation symmetry is broken.

The {\it second term} in Eq. (\ref{eq:model}) is the \emph{bare} Hamiltonian for the acoustic phonons on the honeycomb lattice, which contains the kinetic and elastic energy \cite{Mahan,Shi2012}. %It can be generally expressed as $H_{\mathrm{ph}}=H^{kinetic}_{\mathrm{ph}}+H^{elastic}_{\mathrm{ph}}$~\cite{Mahan,Shi2012}.
% \begin{align}
% H_{\mathrm{ph}}&=H^{kinetic}_{\mathrm{ph}}+H^{elastic}_{\mathrm{ph}}\non\\
% &=\sum_{\kv,i}\frac{\ve{P}_{-\kv}\cdot \ve{P}_{\kv}}{2\rho\delta_V} +\frac{1}{2}\ve{u}_{-\kv}\ve{C}_{\kv} \ve{u}_{\kv}.
% \end{align}
The elastic part can be expressed in terms of the strain tensor $\epsilon_{ij}=\frac{1}{2}(\partial_i u_j+\partial_j u_i)$ and the elastic modulus tensor $C_{ijlk}$, where ${\bf u}=\{u_x,u_y\}$ is the lattice displacement vector.  From symmetry considerations, for a lattice with $C_{6v}$ point group symmetry, there are only two independent non-zero elastic modulus tensor coefficients, $C_{xxxx}$ and $C_{xxyy}$. The phonon Hamiltonian is written in terms of the bosonic operators that appear from quantizing the eigenmodes of the displacement vector ${\bf u}$. We will postpone writing down the elastic energy until next section.

The {\it third term} in Eq.~\eqref{eq:model} denotes the magneto-elastic coupling
that arises from the dependence of the Kitaev coupling $J_K$ on the relative positions between the spins.
Assuming that $J_K$ only depends on the distance $r$ between the atoms and that positions of the spins deviate only slightly from their equilibrium values, we  expand the exchange couplings in powers of the displacement vectors.
Keeping only the lowest order of this expansion, the spin-phonon interaction part of the model Eq.~\eqref{eq:model} can then be written as
\begin{eqnarray}
\label{Cmodel1}
H_{\text c}  &&=- \lambda\sum_{{\bf r},\alpha}  {\bf M}_\alpha
\cdot \left[
{\bf u}({\bf r})- {\bf u}({\bf r}+{\bf M}_\alpha)
\right]  
\sigma_{\bf r}^{\alpha}  \sigma_{{\bf r}+{\bf M}_\alpha}^{\alpha}
\\\nonumber
&&= \lambda\sum_{{\bf r},\alpha}  {\bf M}_\alpha\cdot \left[ \left(
{\bf M}_\alpha \cdot{\bf \nabla} \right)
{\bf u}({\bf r})\right] \sigma_{\bf r}^{\alpha}  \sigma_{ {\bf r}+{\bf M}_\alpha }^{\alpha},
\end{eqnarray}
 where   $\lambda\sim \big(\frac{\diff J_K}{\diff r}\big)_{eq}\ell_a$  characterizes  the strength of the spin-phonon interaction and $\ell_a$ is the lattice constant.
 
%Here, however,  we follow the  symmetry arguments and identify the linear combinations of  the Kitaev interactions that transform as $A_1^{\text sp}$ and $E_2^{\text sp}$ IRRs, and find 
To find the complete set of magneto-elastic couplings, we use symmetry considerations and identify the linear combinations of the Kitaev interactions that transform as $A_1^{\text sp}$ and $E_2^{\text sp}$ irreducible representations (IRRs) of the $C_{6v}$ point group. We find:
\begin{align}
A_1^{\text sp}\sim &(\sigma_{\bf r}^{x}  \sigma_{{\bf r}+{\bf M}_x}^{x}+
\sigma_{\bf r}^{y}  \sigma_{{\bf r}+{\bf M}_y}^{y}+\sigma_{\bf r}^{z}  \sigma_{{\bf r}+{\bf M}_z}^{z})\non\\
E_2^{\text sp} \sim &\{\sigma_{\bf r}^{x}  \sigma_{{\bf r}+{\bf M}_x}^{x}+
\sigma_{\bf r}^{y}  \sigma_{{\bf r}+{\bf M}_y}^{y}-2\sigma_{\bf r}^{z}  \sigma_{{\bf r}+{\bf M}_z}^{z},\non\\
&\,\,\, \sqrt{3}\left(\sigma_{\bf r}^{x}  \sigma_{{\bf r}+{\bf M}_x}^{x}-\sigma_{\bf r}^{y}  \sigma_{{\bf r}+{\bf M}_y}^{y}\right)\}.
\end{align}
Similarly, in the phonon sector, the strain combinations that transform as the IRRs $A_1^{ph}$ and $E_2^{ph}$ are $(\epsilon_{xx}+\epsilon_{yy})$ and $\{\epsilon_{xx}-\epsilon_{yy},2\epsilon_{xy}\}$. Therefore, the spin-phonon couplings that are invariant under $C_{6v}$ consist of two independent channels, one from $A_1^{\mathrm{ph}}\otimes A_1^{\text sp}$ and another one from $E_2^{\mathrm{ph}}\otimes E_2^{\text sp}$, with the coupling constants    $ \lambda_{A_{1}}$ and $ \lambda_{E_{2}}$,  which might be different but of a similar  
strength $\lambda_{A_{1}},\lambda_{E_{2}}\sim \big(\frac{\diff J_K}{\diff r}\big)_{eq}\ell_a\sim J_K$. 
Thus,
the spin-phonon coupling   Hamiltonian  $H_c$   can be written as a sum of two independent contributions invariant under the $C_{6v}$ symmetry, $ H_{\text c} = H_{\text c}^{A_{1}}+H_{\text c}^{E_{2}}$, where
\begin{widetext}
\begin{align}
H_{\text c}^{A_{1}} =&
\lambda_{A_{1}} \sum_{{\bf r}}( \epsilon_{xx}+ \epsilon_{yy} )(\sigma_{\bf r}^{x}  \sigma_{{\bf r}+{\bf M}_x}^{x}+
\sigma_{\bf r}^{y}  \sigma_{{\bf r}+{\bf M}_y}^{y}+\sigma_{\bf r}^{z}  \sigma_{{\bf r}+{\bf M}_z}^{z}),\non\\
H_{\text c}^{E_{2}} =&
\lambda_{E_{2}}\sum_{{\bf r}}[( \epsilon_{xx}- \epsilon_{yy} )(\sigma_{\bf r}^{x}  \sigma_{{\bf r}+{\bf M}_x}^{x}+
\sigma_{\bf r}^{y}  \sigma_{{\bf r}+{\bf M}_y}^{y}-2\sigma_{\bf r}^{z}  \sigma_{{\bf r}+{\bf M}_z}^{z} )+2\sqrt{3}\epsilon_{xy}
(\sigma_{\bf r}^{x}  \sigma_{{\bf r}+{\bf M}_x}^{x}-
\sigma_{\bf r}^{y}  \sigma_{{\bf r}+{\bf M}_y}^{y} )].
\label{Irr-Cmodel}
\end{align}
\end{widetext}
In Sec.\ref{Sec:MFPhcoupling}, we will use the Majorana  fermion representation of the spins to express the spin-phonon coupling $H_c$ in terms of the free Majorana fermion $c_{\ve r}$ and lattice displacement field $\tilde u_{\ve q,\nu}$. %as
%\begin{align}
%H_c=
%\sum_{\ve q,\ve k}
%\psi^T_{-{\bf k}-{\bf q}}(\tilde u_{\ve q,\parallel}\hat \lambda^{\parallel}_{{\bf q},{\bf k}}+\tilde u_{\ve q,\perp}\hat \lambda^{\perp}_{{\bf q},{\bf k}})
%\psi_{\ve k},
%\label{eq:Vertexf}
%\end{align}
%where $\hat \lambda^{\nu}_{{\bf q},{\bf k}}$ are the effective Majorana fermion-phonon (MFPh) coupling vertices.  

%Note that in the Eq.~\eqref{eq:Vertexf} we explicitly  assumed that the $Z_2$ fluxes are not excited by lattice vibrations, i.e.\ the ground state remains in the flux-free sector. 
%note that the simplification of including only Kitaev terms in $H_c$ ensures that the $Z_2$ flux is not excited by lattice strain, i.e., the ground state stays in the flux-free sector. 

%Note that in Eq.~\eqref{eq:Vertexf} we explicitly  assumed that the $Z_2$ fluxes are not excited by lattice vibrations, which follows from the assumption that lattice vibration only changes the strength of Kitaev interaction but doesn't generate other spin interaction terms.
%The explicit  microscopic
%derivation of the  MFPh vertices will be presented in the Sec.\ref{Sec:MFPhcoupling}.

%As we discuss further in the following sections,

\section{Effective action and spectrum for acoustic phonon}  \label{Eltheory}

In order to describe the dynamics of acoustic phonons both in the presence and in the absence of the time reversal symmetry breaking term $\kappa$, it is convenient to move away from the Hamiltonian formulation and employ instead the long-wavelength effective action $\mathcal{S}$ approach. 
Symmetries of the system, such as lattice symmetry and time-reversal symmetry, impose constraints on the number of non-zero independent coefficients in the elastic modulus tensor and viscosity tensor. In the following, only the non-dissipative terms in the effective action are considered.
%The latter is always generated by 
%some kinds of interactions, such as spin-lattice coupling and phonon self-interactions.  
%In the following, only the non-dissipative terms are considered.
%In our case, the non-zero viscosity tensor coefficients appear in the presence of the time-reversal symmetry breaking term $\kappa$ and are generated by the coupling to the bulk Majorana fermion band with non-zero Chern number.\\

%Our analysis follow the ideas in Chapter 18 of  Ref.~\cite{DresselhausBook}.

%%%%%%%%%%%%%%%%%%%%%%%%%

\subsection{Elastic medium with time-reversal symmetry} \label{EltheoryTRS}
We start by considering acoustic phonons in a homogeneous space medium without decay and with time-reversal symmetry.
Their dynamics  can be described  by using the long-wavelength effective action in terms of the 
%displacement
fields $\ve{u}$, which describe the displacement of an atom from its original location.
To lowest order, it reads \cite{LandauElasticity}
\begin{align}
\mathcal{S}_{\mathrm{ph}}^{(s)}&=\int\diff^2 x\diff \tau \, [\rho\, (\partial_{\tau} \ve{u})^2+F^{(s)}],\,\, F^{(s)}=\frac{1}{2}\mathcal{C}_{ijlk}\epsilon_{ij}\epsilon_{lk},
\end{align}
where $\rho$ is the mass density of the lattice ions.  Here the subscript  in $\mathcal{S}_{\mathrm{ph}}^{(s)}$ denotes  that 
%in the absence of the time-reversal symmetry breaking the only non-zero 
the elastic moduli $\mathcal{C}_{ijlk}$ are symmetric under $ij\leftrightarrow lk$, and thus $F^{(s)}$ describes the symmetric part of the elastic energy. 
Due to the symmetry of the strain tensor $\epsilon_{ij}$ under $i\leftrightarrow j$,  the elastic modulus tensor $\mathcal{C}_{ijlk}$ is also symmetric under $i\leftrightarrow j$ and  $l \leftrightarrow k$. Finally, by imposing the $C_{6v}$ lattice symmetry we are left with two independent non-zero elastic modulus tensor coefficients, $C_{xxxx}$ and $C_{xxyy}$.  The elastic energy can then be written as: 
\begin{align}
\label{Appeq:Action0}
F^{(s)}=\big[&C_1 (\epsilon_{xx}+\epsilon_{yy})^2 \\\nonumber & +C_2(\epsilon_{xx}-\epsilon_{yy}+2i\epsilon_{xy})(\epsilon_{xx}-\epsilon_{yy}-2i\epsilon_{xy})\big],
\end{align}
where $C_1 \equiv (C_{xxxx}+C_{xxyy})/2$ and $C_2 \equiv (C_{xxxx}-C_{xxyy})/2$. In momentum and (Matsubara) frequency space, the action is given by
\begin{widetext}
\begin{align}
\mathcal{S}_{\mathrm{ph}}^{(s)}=T\sum_{\ve q,\Omega_n}
\begin{pmatrix}
u_{x,-\ve q} & u_{y,-\ve q}
\end{pmatrix}
\begin{pmatrix}
\rho \,\Omega_n^2+(C_1 +C_2) q_x^2+C_2 q_y^2 & C_1 q_x q_y\\
C_1 q_x q_y & \rho\,\Omega_n^2+(C_1 +C_2) q_y^2+C_2 q_x^2
\end{pmatrix}
\begin{pmatrix}
u_{x,\ve q}\\
u_{y,\ve q}
\end{pmatrix}.
\label{Appeq:PhononAction}
\end{align}
\end{widetext}
We diagonalize Eq.~\eqref{Appeq:PhononAction} through
\begin{align}
\begin{pmatrix}
u_{x,\ve q}\\
u_{y,\ve q}
\end{pmatrix}=
\begin{pmatrix}
\cos\theta_q & -\sin\theta_q \\
\sin\theta_q & \cos\theta_q
\end{pmatrix}
\begin{pmatrix}
\tilde u_{\ve q,\parallel}\\
\tilde u_{\ve q,\perp}
\end{pmatrix},
\label{Appeq:A14}
\end{align}
where $\tilde u_{\ve{q},\nu}$ denotes the magnitude of lattice displacement in the $\nu=\parallel,\perp$ channels which  may be quantized as the acoustic phonons. We then  find the longitudinal and transverse acoustic phonon spectrum and the polarization vectors (defined through $\ve{u}_{\bf q}=\sum_{\nu}\hat{e}_{\ve{q},\nu}\tilde u_{\ve{q},\nu}$)  to be equal to
\begin{align}
\Omega_{\parallel,\ve q}=v_s^{\parallel}q=\sqrt{\frac{C_1+C_2}{\rho}} \,q, \quad  \hat{e}_{\ve{q},\parallel}=\{\cos\theta_q,\sin\theta_q\}\non\\
\Omega_{\perp,\ve q}=v_s^{\perp}q=\sqrt{\frac{C_2}{\rho}} \,q,  \quad \hat{e}_{\ve{q},\perp}=\{-\sin\theta_q,\cos\theta_q\},
\label{appeq:PhSpectrum}
\end{align}
where $q=\sqrt{q_x^2+q_y^2}$  and $\theta_q$ is defined in Fig.\ref{fig:Attenuation} (b). In the case of the bare lattice system, the phonon dispersions are isotropic, and the transverse and longitudinal modes are decoupled, which are generic properties for the 2D medium with $C_{6}$ or higher symmetries.

Another property of elastic media is the  viscosity tensor, which relates the stress tensor, obtained by differentiating the action with respect to the strain tensor,  $\sigma_{ij}=-\frac{\delta \mathcal{S}}{\delta \epsilon_{ij} }$, to the strain rate, 
$ \sigma_{ij}=-\eta_{ijkl}\dot {\epsilon}_{kl}$. The viscosity tensor of non-interacting phonons is identically zero.
However, it becomes finite when the phonon interactions that lead to phonon decay are included. 
In this work, we consider the non-zero viscosity tensor coefficients generated \emph{only} through the spin-lattice coupling. The effects due to, e.g.\ phonon self-interactions, are ignored for simplicity. Moreover, generally, the correction from the phonon self-interactions is subleading in temperature~\cite{Barkeshli2012}.

As the viscosity tensor coefficient coming from the phonon interactions can be anisotropic in space, i.e.\ dependent on $\theta_q$, a straightforward symmetry analysis alone does not allow us to  determine the number of independent components of the symmetric  viscosity tensor $\eta^{(s)}_{ijlk}$ (by the same way as we just did for  $\mathcal{C}_{ijlk}$). 
% because these interactions might be momentum dependent. 
  As such, instead of finding the non-zero $\eta^{(s)}_{ijlk}$ from a symmetry analysis, here we will calculate the decay rate directly from the phonon self-energy by integrating out the Majorana fermions.

%{\cred However, we note that the properties of independent viscosity tensors $\eta_{ijlk}$ doesn't follow the above counting straightforwardly, 
%though the rate of energy decrease in terms of strain-rate tensor $\dot {\epsilon}_{ij}$ and $\eta_{ijlk}$ reads in a similar fashion as
%\begin{align}
%\dot {E}_{ph}=-\frac{1}{2} \eta^{(s)}_{ijlk}\dot{\epsilon}_{ij}\dot{\epsilon_{lk}}.
%\label{eq:action3}
%\end{align}
%This is because $\eta^{(s)}_{ijlk}$ comes from interactions that lead to the phonon decay, which may be momentum dependent, whereas the analysis leading to Eq.~\eqref{SMeq:1-2} assumes momentum independent 4-rank tensor ($\mathcal{C}_{ijlk}$). For concreteness, instead of finding the non-zero $\eta^{(s)}_{ijlk}$ from a symmetry analysis, we calculate the decay rate directly from the phonon self-energy by integrating out the Majorana fermions.}

%%%%%%%%%%%%%%%%%%%%%%%%%

\subsection{Elastic medium with broken time-reversal symmetry}\label{EltheoryTRSbroken}
For an elastic medium with time-reversal symmetry breaking, e.g.\ when $\kappa\neq 0$ in Eq.~\eqref{eq:model}, there is a direct contribution
%a non-trivial correction 
to the phonon effective action coming from the Hall viscosity term~\cite{Avron1995,Barkeshli2012},
\begin{align}
%\mathcal{S}&=\mathcal{S}^{(0)}+\mathcal{S}^{(a)},\nonumber\\
\mathcal{S}_{\mathrm{ph}}^{(a)}=\int\diff^2 x\diff t \, \eta^{(a)}_{ijlk}\epsilon_{ij}\dot{\epsilon_{lk}}, 
\label{eq:HallAction}
\end{align}
 where the viscosity tensor is anti-symmetric, i.e.\  $\eta^{(a)}_{ijlk}=-\eta^{(a)}_{lkij}$. In our case, the time-reversal symmetry breaking comes from the coupling of acoustic phonons with spin excitations, i.e.\  matter Majorana fermions with non-zero Chern number. After integrating out the fermion degrees of freedom, we are left with the phonon effective action $\mathcal{S}_{\mathrm{ph}}^{(a)}$ that breaks time-reversal symmetry. 

The lattice symmetries give additional constraints on the number of independent non-zero Hall viscosity tensor elements. In our case, there is only one independent component $\eta^{(a)}_{xxxy}=\eta_H$. 
The reason is the following: The effective action must be invariant under all symmetry operations, so it must transform as the $A_1$ IRR. As the Hall viscosity tensor is anti-symmetric, we need to look for all the anti-symmetric $A_1$ IRRs formed by the tensor product of two phonon fields, i.e. those satisfying $\eta_{ijlk}=-\eta_{lkij}$. For the elastic medium with $C_{6v}$ symmetry, two phonon fields in the $E_{2}^{\mathrm{ph}}$ irreducible representation (IRR)  can form the tensor product given by $E_{2}^{\mathrm{ph}}\otimes E_{2}^{\mathrm{ph}}\rightarrow A_1^{(s)}+A_2^{(a)}+...$, where the superscript labels if the tensor is symmetric (s) or anti-symmetric (a) under the exchange of the two phonon modes~\cite{DresselhausBook}. With $C_{6v}$ symmetry, there is no anti-symmetric $A_1$ IRR. On the other hand, the time-reversal symmetry breaking term $\kappa$ in the spin Hamiltonian Eq.~\eqref{eq:Kmodel1} lowers the symmetry  from $C_{6v}$ to $C_6$. As a result, the $A_2^{(a)}$ IRR of $C_{6v}$ becomes the $A_1^{(a)}$ IRR of $C_6$, following the compatibility relation~\cite{Koster1963}. Consequently, we obtain one antisymmetric viscosity tensor element.
 
The effective action then becomes 
%also breaks the mirror symmetry in $C_{6v}$. We first impose the the $C_{6v}$ lattice symmetry requires that there  Note that by breaking the time-reversal symmetry, the mirror symmetry in $C_{6v}$ is also broken and $C_{6v}\rightarrow C_6$. The effective action becomes 
\begin{widetext}
\begin{align}
\mathcal{S}_{\mathrm{ph}}^{(a)}=\frac{i \eta_H }{4}\int \diff x^2\diff t \, [(\e_{xx}-\e_{yy}+2i\e_{xy})(\dot\e_{xx}-\dot\e_{yy}-2i\dot\e_{xy})-H.c.]=\int \diff x^2\diff t  \,\eta_H \left[ (\e_{xx}-\e_{yy})\dot \e_{xy}-(\dot \e_{xx}-\dot \e_{yy}) \e_{xy} \right].
\label{Appeq:actiona}
\end{align}
\end{widetext} 
%In order to relate $\eta_H$ with the phonon polarization bubble, in Eq.~\eqref{Appeq:actiona}
$\mathcal{S}_{\mathrm{ph}}^{(a)}$ can also be expressed in terms of the transverse and longitudinal phonon eigenmodes ${\tilde u}_{\ve{q},\nu}$ ($\nu=\parallel,\perp$) as
% the components of the  strain tensor ${\epsilon}_{ij}$   and the components of the strain-rate tensor $\dot {\epsilon}_{ij}$ in terms
 \begin{align}
\mathcal{S}_{\mathrm{ph}}^{(a)}&=\frac{\eta_H}{2}\sum_\qv \int \diff t \, q^2 [\dot{\tilde{u}}_{\qv,\parallel}(t) \tilde{u}_{-\qv,\perp}(t)-\tilde{u}_{\qv,\parallel}(t) \dot{\tilde{u}}_{-\qv,\perp}(t)].
\end{align}
In general, the Hall viscosity coefficient $\eta_H$ can be found from the linear response theory through the Kubo formula, which gives~\cite{Barkeshli2012}
\begin{align}\label{etaH}
\eta_{\mu\nu}(\qv)=\lim_{\Omega\rightarrow 0} \frac{1}{\Omega}\frac{1}{\ell_a^d} \int\diff t \, e^{i\Omega t}\langle [ \frac{\partial H_c}{\partial {\tilde u}_{\qv,\mu}}(t), \frac{\partial H_c}{\partial {\tilde u}_{-\qv, \nu}} (0) ]\rangle,
\end{align}
for an action of the form 
\begin{align}\label{eq:Spha}
\mathcal{S}_{\mathrm{ph}}^{(a)}&=\frac{1}{2}\sum_\qv \int \diff t \, \eta_{\mu\nu}(\qv) \tilde{u}_{-\qv,\mu}(t)\dot{\tilde{u}}_{\qv,\nu}(t),
\end{align}
where we recall that  $H_c$ is the spin-phonon coupling Hamiltonian.

As we will show in Sec.~\ref{Appsec:Hall}, the phonon polarization bubble $\Pi_{\mathrm{ph}}^{\mu\nu}(\qv,\Omega)$ can be expressed in a similar fashion as
\begin{align}
\Pi_{\mathrm{ph}}^{\mu\nu}(\qv,\Omega)=-\frac{i}{2!}\int \diff t\, e^{i\Omega t} \langle T\,  \frac{\partial H_c}{\partial \tilde u_{\qv,\mu}}(t) \, \frac{\partial H_c}{\partial \tilde u_{-\qv, \nu}} (0) \rangle,
\end{align}
Therefore, it is straightforward to find the relation between the response coefficient $\eta_{\mu\nu}(\qv)$ [and  thus $\eta_H= \frac{1}{q^2}\eta_{\perp\parallel}(\qv)$] and $\Pi_{\mathrm{ph}}^{\mu\nu}(\qv,\Omega)$.
%On the other hand, from the linear response theory, one can find the response coefficient in an action of the form 
%\begin{align}\label{eq:Spha}
%\mathcal{S}_{\mathrm{ph}}^{(a)}&=\frac{1}{2}\sum_\qv \int \diff t \, \eta_{\mu\nu}(\qv) \tilde{u}_{-\qv,\mu}(t)\dot{\tilde{u}}_{\qv,\nu}(t),
%\end{align}
% which gives \cite{Barkeshli2012}
%\begin{align}\label{etaH}
%\eta_{\mu\nu}(\qv)=\lim_{\Omega\rightarrow 0} \frac{1}{\Omega}\frac{1}{\ell_a^d} \int\diff t \, e^{i\Omega t}\langle [ \frac{\partial H}{\partial {\tilde u}_{\qv,\mu}}(t), \frac{\partial H}{\partial {\tilde u}_{-\qv, \nu}} (0) ]\rangle.
%\end{align}
%As we will show in Sec.~\ref{Appsec:Hall}, the phonon polarization bubble $\Pi_{\mathrm{ph}}^{\mu\nu}(\qv,\Omega)$ can be expressed in a similar fashion. So it is straightforward to find the relation between the response coefficient $\eta_{\mu\nu}(\qv)$ (and  thus $\eta_H= \frac{1}{q^2}\eta_{\perp\parallel}(\qv)$) and $\Pi_{\mathrm{ph}}^{\mu\nu}(\qv,\Omega)$.
%%%%%%%%%%%%%%%%%%%%%%%%%%%%%%%%%%%%%%%%%%%%%%%%%%%%%%%%%

\section{Microscopic derivation of the effective low-energy coupling Hamiltonian }\label{Sec:MFPhcoupling}

In this section, we express the spin-lattice coupling in terms of the Majorana fermion-phonon (MFPh) coupling. %as the low-energy effective theory and derive the explicit expressions for the MFPh coupling vertices
% $\hat \lambda^{\nu}_{{\bf q},{\bf k}}$ introduced in  Eq~\eqref{eq:Vertexf}.
 To this end, we express the spin operators  in  Eq.~\eqref{Irr-Cmodel} in terms of the Majorana fermions \cite{Kitaev2006} and
obtain the MFPh interaction  in two symmetry channels: 
\begin{widetext}
\begin{align}
H_{\text c}^{A_{1}} =&
-i\lambda_{A_{1}}\sum_{\bf r} ( \e_{xx}+ \e_{yy} )
(c_{{\bf r}, A}c_{{\bf r}+{\bf M}_x, B}+c_{{\bf r}, A}c_{{\bf r}+{\bf M}_y, B}+c_{{\bf r}, A}c_{{\bf r}+{\bf M}_z, B}),\non\\ 
H_{\text c}^{E_{2}} =&
-i\lambda_{E_{2}}\sum_{\bf r}[( \e_{xx}- \e_{yy} )(c_{{\bf r}, A}c_{{\bf r}+{\bf M}_x, B}+c_{{\bf r}, A}c_{{\bf r}+{\bf M}_y, B}-2c_{{\bf r}, A}c_{{\bf r}+{\bf M}_z, B})+2\sqrt{3}\e_{xy}
(c_{{\bf r}, A}c_{{\bf r}+{\bf M}_x, B}-c_{{\bf r}, A}c_{{\bf r}+{\bf M}_y, B} )].
\label{Irr-Cmodel_MF}
\end{align}
\end{widetext}
%Note that from Eq.~\eqref{Irr-Cmodel} to Eq.~\eqref{Irr-Cmodel_MF} is exact because the spin-lattice coupling considered does not excite any flux degrees of freedom.

Since Majorana fermions satisfy  $\{c_{\ve{r},\alpha},c_{\ve{r}',\beta}\}=2\delta_{\ve{r},\ve{r}'}\delta_{\alpha,\beta}$ and
  $\{c_{\ve{k},\alpha},c_{\ve{k}',\beta}\}=\delta_{\ve{k},-\ve{k}'}\delta_{\alpha,\beta}$, where $\alpha,\beta$ denote the sublattice indices A, B, the Fourier transformation of Majorana fermions  is given by 
$c_{{\bf r},A(B)}=\sqrt{\frac{2}{N}}\sum_{\bf k} c_{{\bf k},A(B)}e^{i \ve{k}\cdot \ve{r}}$.   In order to have simpler notations,  in the following we define 
$c_{{\bf k},A}\equiv a_{\bf k}$ and $c_{{\bf k},B}\equiv b_{\bf k}$.
Thus, the MFPh coupling Hamiltonian  in momentum space becomes
\begin{eqnarray}
H_{\text c}=\frac{1}{\sqrt{N}} \sum_{{\bf q},{\bf k}} (H_{{\bf q},{\bf k}}^{A_{1}}+H_{{\bf q},{\bf k}}^{E_{2}})
\end{eqnarray}
 with
 \begin{widetext}
\begin{align}
\label{Irr-Cmodel_MF-q}
H_{{\bf q},{\bf k}}^{A_{1}} =&
-\frac{\lambda_{A_{1}}}{2}(iq_x u_{x,\qv}+ iq_yu_{y,\qv})
\begin{pmatrix}
a_{-{\bf k}-{\bf q}} & b_{-{\bf k}-{\bf q}}
\end{pmatrix}
\left(-f^\prime_{\bf k} {\tau}_y-f^{\prime\prime}_{\bf k} {\tau}_x
 \right)
\left(\begin{array}{c}
a_{{\bf k}} \\b_{{\bf k}}\end{array}\right),\\
H_{{\bf q},{\bf k}}^{E_{2}} =&
-\frac{\lambda_{E_{g}}}{2}(iq_x u_{x,\qv}- iq_yu_{y,\qv})
\begin{pmatrix}
a_{-{\bf k}-{\bf q}} & b_{-{\bf k}-{\bf q}}
\end{pmatrix}
\left(-f^\prime_{1,\bf k} {\tau}_y-f^{\prime\prime}_{1,\bf k} {\tau}_x
 \right)
\left(\begin{array}{c}
a_{{\bf k}} \\b_{{\bf k}}\end{array}\right)\\\nonumber
&-\frac{\lambda_{E_{g}}}{2}(iq_x u_{y,\qv}+ iq_yu_{x,\qv})
\begin{pmatrix}
a_{-{\bf k}-{\bf q}} & b_{-{\bf k}-{\bf q}}
\end{pmatrix}
\left(-f^\prime_{2,\bf k} {\tau}_y-f^{\prime\prime}_{2,\bf k} {\tau}_x
 \right)
\left(\begin{array}{c}
a_{{\bf k}} \\b_{{\bf k}}\end{array}\right).
\end{align}
\end{widetext}
 Here we used the fact that $ \e_{ij} \rightarrow\frac{i}{2}( q_i u_j+q_j u_i) $ and kept  only  leading  in ${\bf q}$ terms.
The auxiliary Pauli matrices $\tau_x$ and  $\tau_y$ in sublattice space are again used in order  to write the   MFPh coupling  in the matrix form. Note that we used a prime (double-prime) to denote the real (imaginary) part. We also defined
\begin{align}
f_{\bf k}&=2J\left( e^{i{\bf k}\cdot{\bf n}_1} + e^{i{\bf k}\cdot{\bf n}_2} +1  \right), \\
f_{1,{\bf k}}&=2J\left( e^{i{\bf k}\cdot{\bf n}_1} + e^{i{\bf k}\cdot{\bf n}_2} -2  \right), \\
f_{2,{\bf k}}&=2\sqrt{3}J\left( e^{i{\bf k}\cdot{\bf n}_1} - e^{i{\bf k}\cdot{\bf n}_2}\right).
\label{f-sigma}
\end{align}
In order to obtain the MFPh coupling vertices, we express the phonon modes
in terms of the transverse and longitudinal eigenmodes defined in 
Eq.(\ref{Appeq:A14}).
% We find
%\begin{align} 
%i(q_x u_{x,\qv} + q_y u_{y,\qv})&=i q \,\tilde u_{q,\parallel},\nonumber\\  
%i(q_x u_{x,\qv} - q_y u_{y,\qv})&=\frac{i}{q}\left[ (q_x^2-q_y^2)\tilde u_{q,\parallel}-2q_xq_y \tilde u_{q,\perp}\right]
%,\nonumber\\
%i(q_x u_{y,\qv} + q_y u_{x,\qv})&=\frac{i}{q}\left[ 2q_xq_y \tilde u_{q,\parallel}+(q_x^2-q_y^2)\tilde u_{q,\perp}\right].
%\end{align}
This gives
\begin{align}
\label{eq:Vertices}
H_{{\bf q},{\bf k}}^{\parallel} =& \tilde u_{\qv,\parallel}
\begin{pmatrix}
a_{-{\bf k}-{\bf q}} & b_{-{\bf k}-{\bf q}}
\end{pmatrix}
\hat \lambda^{\parallel}_{{\bf q},{\bf k}}
\left(\begin{array}{c}
a_{{\bf k}} \\b_{{\bf k}}\end{array}\right),
\nonumber\\
H_{{\bf q},{\bf k}}^{\perp} =& \tilde u_{\qv,\perp}
\begin{pmatrix}
a_{-{\bf k}-{\bf q}} & b_{-{\bf k}-{\bf q}}
\end{pmatrix}
\hat \lambda^{\perp}_{{\bf q},{\bf k}}
\left(\begin{array}{c}
a_{{\bf k}} \\b_{{\bf k}}\end{array}\right),
\end{align}
where the   MFPh vertices are 
\begin{widetext}
\begin{align}
\label{eq:lambdas}
\hat \lambda^{\parallel}_{{\bf q},{\bf k}}=&\frac{i\lambda_{A_{1}}}{2} q 
\left(f^\prime_{\bf k} {\tau}_y+f^{\prime\prime}_{\bf k} {\tau}_x
 \right)+\frac{i\lambda_{E_{2}}}{2} q \left[
 \cos 2\theta_q
\left(f^\prime_{1,\bf k} {\tau}_y+f^{\prime\prime}_{1,\bf k} {\tau}_x
 \right)+\sin 2\theta_q
\left(f^\prime_{2,\bf k} {\tau}_y+f^{\prime\prime}_{2,\bf k} {\tau}_x
 \right)
\right],\nonumber\\
\hat \lambda^{\perp}_{{\bf q},{\bf k}}=&\frac{i\lambda_{E_{2}}}{2} q \left[
- \sin 2\theta_q
\left(f^\prime_{1,\bf k} {\tau}_y+f^{\prime\prime}_{1,\bf k} {\tau}_x
 \right)+\cos 2\theta_q
\left(f^\prime_{2,\bf k} {\tau}_y+f^{\prime\prime}_{2,\bf k} {\tau}_x
 \right)
\right].
\end{align}
\end{widetext}
%Expanding   $\hat \lambda^{\parallel}_{{\bf q},{\bf k}}$ and $\hat \lambda^{\perp}_{{\bf q},{\bf k}}$ near the Dirac points $\pm K$ of the Majorana fermion spectrum, we obtain the Eq.~(\ref{eq:VertexLowE}) in the main text.
%%%%%%%%%%%%%%%%%%%%%%%%%%%%%%%%%%%%%%%%%%%%%%%%%%%%%%%%%
To the leading order in momentum $q$ and near the Dirac points  $\pm K$ of the Majorana fermion spectrum, the MFPh vertices $\hat \lambda^{\nu}_{{\bf q},{\bf k}}$ are given by
\begin{align}
\hat \lambda^{\parallel}_{{\bf q},\pm K+{\bf k}}\equiv\hat \lambda^{\parallel}_{{\bf q},{\bf k}}=& 3\,i\, q\,\lambda_{E_2}  (\pm \sin 2\theta_q\tau^x-\cos 2\theta_q\tau^y),\non\\
\hat \lambda^{\perp}_{{\bf q},\pm K+{\bf k}}\equiv\hat \lambda^{\perp}_{{\bf q},{\bf k}}=& 3\,i\, q\,\lambda_{E_2}  (\pm\cos 2\theta_q\tau^x+\sin 2\theta_q\tau^y),
\label{eq:VertexLowE}
\end{align}
where now ${\bf k}$ denotes the deviation from the Dirac point.
Since the MFPh coupling in the $A_1$ channel  appears at  higher orders in $\ve k$, whenever the computation is restricted to the low-energy Majorana fermions near the $\pm K$ points, the $A_1$ channel is ignored and only 
the dominant contribution in the $E_2$ channel is considered.

%is ignored, because it is found to be at higher order in $\ve k$, and thus only contributes at higher order in temperature to $\alpha_s$. 
%Note that the coupling in the $A_1$ channel is ignored, because it is found to be at higher order in $\ve k$, and thus only contributes at higher order in temperature to $\alpha_s$. 

\section{Phonon Propagator}\label{SMphonons}
Our next task is to integrate out the Majorana fermions and obtain an effective low-energy theory only for the phonons.

%%%%%%%%%%%%%%%%%%%%%%%%%%%%%
\subsection{Free phonon propagator}\label{BarePhonon}
The free phonon propagator in terms of lattice displacement field $\tilde{u}_{\ve{q},\nu}$  defined in Eq.\eqref{Appeq:A14} is given by
\begin{align}
D_{\nu\nu,{\bf q}}^{(0)}(t)=-i\langle T \tilde{u}_{-{\bf q},\nu}(t)\tilde{u}_{{\bf q},\nu}(0)\rangle^{(0)},
\label{BarePHpropagator}
\end{align}
where the superscript $(0)$ denotes the bare propagator, and $\nu=\parallel,\perp$ labels the polarization. 
%We follow the convention in Ashcroft$\&$Mermin~\cite{ashcroft1976solid}, the second quantized form of a displacement field is
%\begin{align}
%{{\ve u}}_{\nu}(x,t)=i\frac{1}{\sqrt{N}}\sum_{\bf q} \big(\frac{\hbar}{2\rho\,\delta_V \Omega_{\bf q}}\big)^{1/2} \hat{e}_{{\bf q},\nu}(a_{\bf q} e^{-i\Omega_{\bf q} t}+a^{\dg}_{-{\bf q}}e^{i\Omega_{\bf q} t})e^{i{\bf q}\cdot x},
%\end{align}
% and 
 The second quantized form of $ \tilde{u}_{{\bf q},\nu}(t)$ is given by \cite{ashcroft1976solid}
\begin{align}
 \tilde{u}_{{\bf q},\nu}(t)=i\big(\frac{\hbar}{2\rho\,\delta_V \Omega_{\bf q}}\big)^{1/2} (a_{\bf q} e^{-i\Omega_{\bf q} t}+a^{\dg}_{-{\bf q}}e^{i\Omega_{\bf q} t}),
\end{align}
where $\delta_V$ is the area enclosed in one unit cell and $\rho$ is the mass density of the lattice ions. 
The time-ordered phonon propagator in the momentum and frequency space is then given by
\begin{align}
D_{\nu\nu}^{(0)}({\bf q},\Omega)=\int\diff t D^{(0)}_{\nu\nu,{\bf q}}(t)e^{i\Omega t}=-\frac{\hbar}{\rho\,\delta_V}\frac{1}{\Omega^2-\Omega_{\bf q}^2+i\delta}.
\end{align}
In the rest of the discussions, we set $\hbar=1$. \\

%%%%%%%%%%%%%%%%%%%%%%%%%%%%%%%%%%%%%%%%%%%%%%%%%%%%%%%%%

%%%%%%%%%%%%%%%%%%%%%%%%%%%%%%%%%%%%%%%%%%%%%%%%%%%%%%%%%
\subsection{Phonon polarization bubble}
\label{Appsec:Polarization}

To compute the corrections to the effective phonon action due to the spin-lattice coupling, we calculate the phonon one-loop self-energy  shown in Fig.\ref{fig:Attenuation} (b) that can be expressed as
\begin{widetext}
\begin{align}
\Pi_{\mathrm{ph}}^{\mu\nu}(\qv,\Omega)
&=\frac{-i}{2!}\int \diff t e^{i\Omega t} \frac{1}{N} \sum_{\kv,\kv'}\langle T\,  \psi_{-\kv-\qv}^T(t) \hat\lambda_{\qv,\kv}^{\mu}\psi_\kv(t) \psi_{-\kv'+\qv}^T(0) \hat\lambda_{-\qv,\kv'}^{\nu}\psi_{\kv'}(0)\rangle\non\\
&=i\tr{[\hat\lambda^{\nu}_{\qv, \kv}\mathcal{G}(\kv,\omega)\hat\lambda^{\mu}_{ \qv, \kv}\mathcal{G}(\kv-\qv,\omega+\Omega)]},
\label{Appeq:C10}
\end{align}
\end{widetext}
where  $\mu,\nu=\parallel,\perp$  and $\mathcal{G}(\kv,\omega)$ denotes the Majorana fermions Green's function given by
\begin{widetext}
\begin{align}
\mathcal{G}({\bf k},\omega)&=-i\int_{-\infty}^{+\infty}\diff t\langle T \psi_{\bf k}(t) \psi_{-{\bf k}}^T(0)\rangle e^{i\omega t}=\frac{1}{2}\left[ 
\frac{1}{\omega+\df{\kv}-i\delta}\left(I-\frac{\vec{\mathcal{F}}_{\bf k}\cdot\vec{\tau}}{\df{\kv}}\right)+\frac{1}{\omega-\df{\kv}+i\delta}\left(I+\frac{\vec{\mathcal{F}}_{\bf k}\cdot\vec{\tau}}{\df{{\bf k}}}\right)\right]\non\\ 
&=\frac{1}{2}\sum_{s=\pm}\,\frac{1}{\omega+s\df{{\bf k}}-s\,i\delta}\left[I-s\frac{\vec{\mathcal{F}}_{\bf k}\cdot\vec{\tau}}{\df{{\bf k}}}\right],
\label{MFGF}
\end{align}
\end{widetext}
where $\vec{\mathcal{F}}_{\bf k}=\{-\im f_{\bf k},-\re f_{\bf k},\Delta_{\bf k}\}$ [see App.~\ref{MajoranaGF} for details of the derivation]. Recall that $\Delta_{\bf k}=4\kappa \big(\sin \ve k\cdot\ve n_1-\sin \ve k\cdot\ve n_2+\sin \ve k\cdot (\ve n_1-\ve n_2)\big)$ is only non-zero when time-reversal symmetry is broken.
We  also use $\hat\lambda_{\qv,\kv-\qv}=\hat\lambda_{\qv,\kv}+\mathcal{O}(q^2)$, and drop the subleading $\mathcal{O}(q^2)$ terms. 
$\tr[...]$ in Eq.~\eqref{Appeq:C10}  sums over momentum ${\bf k}$, sublattice degrees of freedom (A and B) and integrates over frequency as $\int \frac{\diff \omega}{2\pi}$. 
%The arrows in $\overleftrightarrow{D}, \overleftrightarrow{\Pi}$ indicate the matrix form of the phonon Green's function and self-energy.
  Note that since the Majorana fermions  are real,  the Majorana fermionic excitation at $({\bf k},\omega)$ and $(-{\bf k},-\omega)$ are physically the same, so  $\mathcal{G}({\bf k},\omega)=-\mathcal{G}(-{\bf k},-\omega)^T$. We also note that since our computation is performed directly in terms of Majorana fermions, all ${\bf k}$ and $\omega$ modes should be summed over, which is different from the calculations that are performed  in terms of the complex-fermion representation.
%The Majorana fermion feature is taken care of when we define the Fourier transformation in Eq.~\eqref{eq:Fourier}.

The renormalized phonon propagator is then given by the equation
\begin{eqnarray}\label{renphpropagator}
D(\qv,\Omega)=\left\{
\left(D^{(0)}(\qv,\Omega)\right)^{-1}-\Pi_{\mathrm{ph}}(\qv,\Omega)
\right\}^{-1}.
\end{eqnarray}
For calculations at finite temperature $T$,  it is convenient to express the Majorana fermions Green's function and the phonon polarization bubble in  the Matsubara frequency representation: 
\begin{align}
\label{eq:MFGreenFunc}
\mathcal{G}(\kv,i\omega_m)&=\frac{1}{2}\sum_{s=\pm}\,\frac{1}{i\omega_m+s|\vec{\mathcal{F}}_\kv |} \left( I-s\frac{\vec{\mathcal{F}}_\kv\cdot\vec{\tau}}{\df{\kv}} \right)
\end{align}
and
\begin{align}\label{Appeq:Polarization}
\Pi_{\mathrm{ph}}^{\mu\nu}&(\qv,i\Omega_n)=\non \\
-&\tr [ \hat\lambda^{\nu}_{ \qv, \kv}\mathcal{G}(\kv,i\omega_m)\hat\lambda^{\mu}_{ \qv, \kv}\mathcal{G}(\kv-\qv,i(\omega_m+\Omega_n))],
\end{align}
where $\tr[...]$ now sums over momentum $\kv$, Matsubara frequencies $i\omega_m$ as $T\sum_m$ and sublattice degrees of freedom.  
In general, the phonon polarization bubble $ \Pi_{\mathrm{ph}}$ contributes  to  the renormalization of the sound velocity~\cite{Mahan}, mixing of the transverse and longitudinal phonon modes~\cite{Barkeshli2012}, and attenuation of sound waves~\cite{Plee2011,Plee2013}.  We compute  these effects explicitly  in Secs.~\ref{sec:Attenuation} and~\ref{Sec:phonTRSbr}.

As a final remark, we note that the corrections to the spin excitation dynamics from the spin-lattice coupling are relatively weak. The Majorana fermion self-energy [Fig.\ref{fig:Attenuation} (c)] can be expressed as
\begin{align}
\Sigma({\bf k},\omega)&=i\tr{[\hat\lambda^{\nu}_{ {\bf q}, {\bf k}}D_{\nu\nu}^{(0)}({\bf q},\Omega)\hat\lambda^{\nu}_{ {\bf q}, {\bf k}}\mathcal{G}({\bf k}-{\bf q},\omega-\Omega)]},
\label{Appeq:MFSE}
\end{align}
where the bare  phonon propagator in terms of the lattice displacement field $\tilde{u}_{\ve{q},\nu}$ is given by  Eq.(\ref{BarePHpropagator}).
   Here again we only keep terms linear in ${\ve q}$ and ${\ve k}$ in the MFPh coupling vertices, so  
 $\hat\lambda_{{\bf q},{\bf k}-{\bf q}}\simeq \hat\lambda_{{\bf q},{\bf k}}$. 
 In addition, $\tr[...]$ sums over momentum ${\ve q}$ and polarization $\nu$ and integrates over the phonon frequency as $\int \frac{\diff \Omega}{2\pi}$. From Eq.~\eqref{Appeq:MFSE}, we find that the imaginary part of $\Sigma \sim T^2$, which is smaller than the typical fermion energy $\sim T$.

% ($k$ and $-k$), frequency ($\omega$ and $-\omega$) and sublattice degrees of freedom (A and B). %As the $Z_2$ flux excitation in pure Kitaev model is static, the gauge fluctuations do not enter into the correction to the phonon propagator at $T=0$, and is suppressed exponentially at low enough temperature $T\ll \Delta_{flux}$. 
%\new{We also note that our computation is performed directly in terms of Majorana fermions, so the Majorana fermion Green's function with both $k,\,-k$ and $\omega, -\omega$ should be summed up. }

%\begin{align}
%&\mathcal{G}( k,\omega)=-i\langle 
%\psi_{ k}\psi_{- k}^T
%\rangle_{\omega}=\non\\
%&\frac{1}{2}\{\frac{1}{\omega+|f_k|}(I-\frac{\vec{\mathcal{F}}_k\cdot\vec{\tau}}{|f_k|})+\frac{1}{\omega-|f_k|}(I+\frac{\vec{\mathcal{F}}_k\cdot\vec{\tau}}{|f_k|})\},
%\label{eq:MfG}
%\end{align}
%where $\vec{\mathcal{F}}_k=\{-\im f_k,-\re f_k,0\}$.
%Eq.~\eqref{} allows us to obtain the phonon self-energy due to Majorana fermion fluctuations diagrammatically in the same way as due to electrons/complex fermionic spinons.

%explicitly the sound attenuation ($\Pi^{\nu\nu}_{ph}$) and the mixing of transverse and longitudinal modes ($\Pi^{\perp\parallel}_{ph}$) in the following.

%%%%%%%%%%%%%%%%%%%%%%%%%%%%%%%%%%%%%%%%%%%%%%%%%%%%%%%%%

\section{Analytical calculation of attenuation coefficient} \label{sec:Attenuation}

The quantitative description of the  attenuation process can be obtained  
through the lossy acoustic wave function  which  
decays with distance away from the driving source as
\begin{align}
\label{eq:WaveFunc}
\ve u(\ve x,t)=\ve u_0 e^{-\alpha_s (\ve q) x}e^{i(\Omega t- \ve q\cdot \ve x)}.
\end{align}
where ${\ve u}(\ve x,t)$ is the  lattice displacement vector, $\ve u_0 = {\ve u}(\ve x,t=0)$, $\Omega$  is the acoustic wave frequency and $\ve q$ is the propagation vector. The attenuation coefficient $\alpha_s (\ve q)$, defined as the inverse of the phonon mean free path, can be  calculated from the imaginary part of the  phonon self-energy  as  
\begin{align}
%\alpha^{\nu}_s=-\frac{2}{v_{\nu}}\im [\Pi^{\nu\nu}_{ph}(\ve q,\Omega)]_{\Omega=v_{\nu}|\ve q|}, \text{ where } \nu=\parallel,\perp.
\alpha_s(\ve q)\propto-\frac{1}{v^2_s q}\im [\Pi_{\mathrm{ph}}(\ve q,\Omega)]_{\Omega=v_s q},
\label{eq:Attenuation}
\end{align}
where $v_s$ is the sound velocity. The derivation of this result is shown in Appendix \ref{Appsec:alphas}.

\subsection{Sound attenuation coefficient at $T>0$}\label{attenT}

The lowest-order Majorana fermion-phonon interaction leading to the phonon damping  is shown diagrammatically in Fig.~\ref{fig:Attenuation}(b) with $\mu=\nu$. In this section, we will evaluate the sound attenuation coefficient $\alpha_s (q)$   by relating it to the imaginary part of the diagonal components of the polarization bubble $\im \Pi_{\mathrm{ph}}^{\nu\nu}(\qv,\Omega)$ defined in Eq.(\ref{Appeq:C10}). 
We  will consider the sound attenuation in the temperature range $v_F q  < T < \Delta_{\mathrm{flux}}$, where $v_F$ is the Majorana fermion velocity at the Dirac point, and  assume $v_s<v_F$, the situation which can be potentially realized in Kitaev materials~\cite{Winter2016,Hirobe2017}.
Under these assumptions, there is a finite phase space for the scattering of phonons with thermally excited low-energy Majorana fermions [see Fig.~\ref{fig:Attenuation} (a)]. 
%Under these assumptions, the scattering of phonons  comes from the on-shell processes involving Majorana fermions  and has a  finite  phase space. 
 
We first note that from the kinematic constraints, $\im \Pi_{ph}^{\nu\nu}(q,\Omega)\neq 0$ only at $T>0$ when $v_s<v_F$.  
To compute $ \Pi_{\mathrm{ph}}^{\nu\nu}(\qv,\Omega)$ at finite temperature, it is convenient to first evaluate in the Matsubara frequency representation, i.e.\ $ \Pi_{\mathrm{ph}}^{\nu\nu}(\qv,i\Omega_n)$ in Eq.~\eqref{Appeq:Polarization}, and then perform the analytical continuation to the real frequency.
Note that dynamical part only appears in the denominator of the Majorana fermion Green's function [see Eq.~\eqref{eq:MFGreenFunc}]. Thus, we can first evaluate the frequency summation~\cite{AltlandBook}:
% the time-ordered response functions~\cite{AltlandBook}
\begin{widetext}
\begin{align}
\mathbb{P}^{s,s'}(\kv,\qv,i\Omega_n)&
\equiv T\sum_m \frac{1}{i\omega_m+s|\nf{\kv}|}\frac{1}{i(\omega_m+\Omega_n)+s'|\nf{\kv-\qv}|}=\frac{1}{2}\frac{\big(s\tanh{\beta|\nf{\kv}|/2}-s'\tanh{\beta|\nf{\kv-\qv}|/2}\big)}{i\Omega_n+s'|\nf{\kv-\qv}|-s|\nf{\kv}|},
\label{eq:D8}
\end{align}
\end{widetext}
where $s,s'$ correspond to  the $\pm$ signs in the Majorana fermion Green's function $\mathcal{G}(\kv,i\omega_m)$ [see Eq.~\eqref{eq:MFGreenFunc}]. Performing  the analytical continuation to the real frequency, we get
\begin{align}
\mathbb{P}^{s,s'}(\kv,\qv,\Omega)&=\frac{1}{2}\frac{\big(s\tanh{\beta|\nf{\kv}|/2}-s'\tanh{\beta|\nf{\kv-\qv}|/2}\big)}{\Omega+s'|\nf{\kv-\qv}|-s|\nf{\kv}|+i\delta \sgn \Omega}.
\label{eq:D9}
\end{align}
%where the $i\delta$ prescription follows that for the time-ordered Green's function of a boson field.
%where we have specified the $i\delta$ prescription for the time-ordered response function {\cred $\mathbb{P}^{s,s'}(k,q,\Omega)$}. 
Note that $\im\mathbb{P}^{s,s'}(\kv,\qv,\Omega)\neq 0$ only when there is a pole in the denominator of Eq.~\eqref{eq:D9}, i.e.\ $\Omega+s'|\nf{\kv-\qv}|-s|\nf{\kv}|=0$, which for $v_s<v_F$  requires  the same choice of the sign in $s$ and $s'$, i.e.\ $s s'>0$. Moreover, when $s s'>0$, the numerator of Eq.~\eqref{eq:D9} is finite only when $|\nf{\kv}|, |\nf{\kv-\qv}|\lesssim T$, and is exponentially suppressed when $|\nf{\kv}|, |\nf{\kv-\qv}|\gtrsim 1/\beta\sim T$. 
 
In the temperature range of our interest,  $T< \Delta_{\mathrm{flux}}\ll J_K$, the spectrum of the Majorana fermions can be  linearized near  the Dirac points $\pm K$,  and  the MFPh vertices $\hat \lambda^{\nu}_{{\bf q},{\bf k}}$ are given by Eq.~\eqref{eq:VertexLowE}.
%Because the linear approximation near the Dirac points should break down only at a high energy scale of $J_K$, which is much larger than the temperature $T$.
In this approximation, the MFPh vertices are constant in terms of $\ve k$  and thus the momentum summation in Eq.~\eqref{Appeq:Polarization} can be obtained by first  replacing the true MFPh vertices with the Pauli matrices $\tau^{x,y}$. We then get
\begin{align}
&\tilde \Pi_{\alpha\beta}(\qv,\Omega)\equiv\tilde \Pi_{\alpha\beta}=\non\\
&\tr{[\tau^{\alpha}\mathcal{G}(\kv,i\omega_m)\tau^{\beta}\mathcal{G}(\kv-\qv,i(\omega_m+\Omega_n))]}_{i\Omega_n\rightarrow \Omega+i\delta},
\label{eq:Polarizationxy}
\end{align}
where $\alpha,\beta=x,y$.  The explicit expressions for $\tilde \Pi_{xx}$, $\tilde \Pi_{yy}$ and $\tilde \Pi_{xy}$ are derived in the Appendix \ref{Appsec:Attenuation}.

The diagonal components of  $\im \Pi_{\mathrm{ph}}^{\nu\nu}(\qv,\Omega)$ can now be expressed 
in a straightforward way in terms of $\tilde \Pi_{\alpha\beta}$ as
\begin{widetext}
\begin{align}
\Pi_{\mathrm{ph}}^{\parallel\parallel}(\qv,\Omega)&=9\lambda_{E_2}^2q^2[\sin^2 2\theta_q\tilde \Pi_{xx}+\cos^2 2\theta_q\tilde \Pi_{yy}-\sin 2\theta_q \cos 2\theta_q(\tilde \Pi_{xy}+\tilde \Pi_{yx})],\label{Piparallel}\\
\Pi_{\mathrm{ph}}^{\perp\perp}(\qv,\Omega)&=9\lambda_{E_2}^2q^2[\cos^2 2\theta_q\tilde \Pi_{xx}+\sin^2 2\theta_q\tilde \Pi_{yy}+\sin 2\theta_q \cos 2\theta_q(\tilde \Pi_{xy}+\tilde \Pi_{yx})].\label{Piperp}
\end{align}
\end{widetext}
In the presence of time-reversal symmetry, i.e.\ when $\kappa=0$, which corresponds to the $z$ component of $\nf{\kv}$ vanishing, we find that the imaginary part of the phonon polarization bubble for the longitudinal and transverse polarizations are (see Appendix \ref{Appsec:Attenuation} for derivation):
\begin{align}
\label{PolarizationL}
\im \Pi_{\mathrm{ph}}^{\parallel\parallel}(\qv,\Omega)
\approx& -\frac{36\pi\lambda_{E_2}^2 q |\Omega|}{v_F^3\mathcal{A}_{\text{BZ}}}T (1-\cos 6\theta_q) \ln 2,\\
\im \Pi_{\mathrm{ph}}^{\perp\perp}(\qv,\Omega)\approx&-\frac{36\pi\lambda_{E_2}^2 q |\Omega|}{v_F^3\mathcal{A}_{\text{BZ}}}T (1+\cos 6\theta_q) \ln 2.
\label{PolarizationT}
\end{align}
In a system with time-reversal-symmetry breaking, from Eqs.~\eqref{eq:Kmodels} and~\eqref{eq:Kmodelf}, we note that the $\kappa$ term does not excite $Z_2$ flux but opens a gap in the Majorana fermion bands with Chern number $\pm 1$~\cite{Kitaev2006}. The correction to the sound attenuation is perturbative in $\kappa$. Moreover, when the gap at $\pm K$ is small, i.e.\ $\kappa<v_F q$, the correction is negligible. We do not consider this correction to the sound attenuation further.
%i.e. in  a magnetic field, we note that as $\Delta_{ K}=6\sqrt{3}\kappa\sim h^3/J_K^2$, the correction to the sound attenuation is small  when $h^3/J_K^2\lesssim v_F q/2$. We do not consider this correction further.

To understand how the constraint $v_F q<T<\Delta_{\mathrm{flux}}$ is imposed in the computation, we first note that when $v_s<v_F$, the thermally excited Majorana fermions that can scatter off of the acoustic phonon should have minimum energy $\sim v_F q/2$. To optimize the phase space for phonon decay, we require $v_F q\ll T$. On the other hand, when $T\sim \Delta_{\mathrm{flux}}$, the thermally excited flux degrees of freedom should be important, and  the zero flux sector approximation breaks down. These processes remain to be considered  in a future work.
%Having computed $\im \Pi^{\parallel\parallel}(\qv,\Omega)$ and $\im \Pi^{\perp\perp}(\qv,\Omega)$, we have now all the information about the sound attenuation coefficient $\alpha^\nu_s$
%in the transverse and the longitudinal channels.
%The attenuation coefficient for transverse/longitudinal modes are in general different depending on the corresponding sound velocity $v^{\nu}_s$ and MFPh vertex.   
%\new{We consider the limit $v_F>v_s$ in the following analysis, which might be true for Kitaev material $\alpha$-RuCl$_3$\footnote{Estimating the sound velocity $v_s\sim 4.4\times 10^3 m/s$~\cite{Widmann2019} and Majorana fermion velocity $v_{F}\sim 3J_K\ell_a\sim 1.4\times 10^4 m/s$~\cite{Winter2016},where $\ell_a$ is the average lattice constant, we have $v_{F}> v_s$.}.} 
%As we discusses above, in the limit $v_s<v_F$, the imaginary part of phonon self-energy  and thus the attenuation coefficient are non-zero only at $T>0$, and the phonon decay is due to  the physical on-shell processes  shown in Fig.~\ref{fig:Attenuation} (a). 
%Namely, only the low-energy Majorana fermion modes near the Dirac points are involved in the scattering with acoustic phonons, the minimum energy of Majorana fermion is $\sim v_F q/2$. To optimize the phase space for phonon decay, we require $v_F q\ll T$. 
%In this low-energy limit, from Eqs.~\eqref{PolarizationL}, and ~\eqref{PolarizationT}, we find $\im [\Pi^{\nu\nu}_{\mathrm{ph}}(\ve q, \Omega=v_s^{\nu}q)]\varpropto q|\Omega|T$. Thus, from Eq.~\eqref{eq:Attenuation} we obtain

From Eq.~\eqref{SMeq:1-5}, we find the sound attenuation coefficient to be equal to
\begin{align}
\alpha_s^{\nu}(\ve q)\sim \big(\frac{\lambda_{E_2}}{v_F/\ell_a}\big)^2\frac{v_s^{\nu}}{v_F}\frac{T}{C_{\nu}\delta_V}q (1\mp\cos 6\theta_q),
\label{eq:13}
\end{align} 
% ${\cred \gamma_{\alpha}}=0.41$
where $C_{\nu}\delta_V=m_{ion}(v_s^{\nu})^2$ is determined by the elastic modulus tensor coefficient of the material, where $C_{\parallel}=C_1+C_2$ and $C_{\perp}=C_2$, and the upper/lower signs are for $\nu=\parallel$ and $\nu=\perp$ polarizations, respectively.  Importantly, $(1\mp \cos6\theta_q)$ is highly anisotropic and vanishes along certain directions, as shown in Fig. \ref{fig:summary}(b). We note that the $\alpha_s$ due to the phonon-Majorana fermion coupling should be much larger than that from the phonon anharmonic interaction. As $\frac{\lambda_{E_2}}{v_F/\ell_a}\sim 1,\, \frac{v_s^{\nu}}{v_F}\sim 1$, the main suppression of $\alpha_s$ comes from the ratio $T/(C_\nu \delta V)$, where $(C_\nu \delta V)$ should be a large energy of order $eV$. On the other hand, as $(C_\nu \delta V)$ appears for a generic sound attenuation mechanism, at low T, $\alpha_{s,ph-ph}/\alpha_{s,ph-f}\sim \big(\frac{T}{v_s/\ell_a}\big)^2\ll 1$, so the sound attenuation due to Majorana fermion-phonon coupling is dominant.

 We note two important features of the attenuation coefficient. \emph{First}, $\alpha_s$ scales linearly with temperature, which counts the phase space of the Majorana fermion scattering. %Estimating the $3D$ bulk elastic modulus as $\sim 10GPa$~\cite{}, the elastic energy is $\sim 10eV$. A rough estimate gives $\alpha_s\sim q T/(100eV)$ assuming a large magneto-elastic coupling such that $\lambda_{E_2}/(v_F\ell_a)\sim 1$~\cite{KimKee2016}. 
\emph{Second}, while the phonon spectrum is isotropic at the leading $q^2$ order due to the six-fold rotation symmetry of the lattice, $\im [\Pi^{\nu\nu}_{\mathrm{ph}}(\ve q, \Omega)]$ can be anisotropic due to the coupling to the fermions with the \emph{Dirac} spectrum. The angular modulation factor $\cos 6\theta_q$ may be understood from symmetry considerations as follows. The phonon polarization bubble [see Eq.~\eqref{Appeq:C10}] is a convolution of two  MFPh coupling vertices (each in the $E_2^{ph}\otimes E_2^{sp}$ representation) and two Majorana fermion propagators [each $\mathcal{G}$ near the Dirac points in Eq.~\eqref{MFGF} contains $E_1^{MF}$ IRR]. So $\alpha_s$, in the identity IRR ($A_1$), comes as the product of $E_{2}^{ph}\otimes E_{2}^{sp}\otimes E_{1}^{MF}\otimes E_{2}^{ph} \otimes E_{2}^{sp}\otimes E_{1}^{MF}$ in Eq.~\eqref{Appeq:C10}, which  gives an angular modulation $\cos 6\theta_q$.

%In passing, we comment on a few 
%To go beyond the pure Kitaev model, we note that
%For simplicity, we consider the model with $C_{6v}$ symmetry in zero field. However, the star-like feature of sound attenuation rate should be qualitatively preserved with weak lattice anisotropy.  
%\MY{MY: Add a discussion on the physical picture of sound attenuation}

%%%%%%%%%%%%%%%%%%%%%%%%%%%%%%%%%%%%%%%%%%%%%%%%%%%%%%%%%

\section{Phonon dynamics with time-reversal-symmetry breaking}
\label{Sec:phonTRSbr}

In this section, we will study the observable consequences of the Majorana fermion-phonon interaction in the presence of time reversal symmetry breaking, for example, due to an applied magnetic field. 
As was discussed above, 
the leading order perturbation from the magnetic field is the three-spin interaction  $\kappa$ term in the spin Hamiltonian, which does not change the sound attenuation coefficient qualitatively. On the other hand, the phonon system experiences Berry curvature induced by the $\kappa$ term due to the spin-lattice coupling. In terms of the effective action, this is the Hall viscosity term given by Eq.~\eqref{eq:HallAction}. 
%The Hall viscosity may lead to a few observable consequences, such as the intrinsic phonon thermal Hall conductivity~\cite{Shi2012,Rosch2018}.
Here, we first compute  the Hall viscosity coefficient by relating it to the off-diagonal component of the polarization bubble that is odd under time reversal, and then show  how it renormalizes the phonon spectrum. \\
  % Both the perturbative and non-perturbative corrections in terms of $\kappa$ are obtained.  
 % We show that it also renormalizes the phonon spectrum in an unusual way.

%In this section, we  will study the observable consequences
%of  the Majorana fermion-phonon interaction in the presence of time reversal symmetry breaking, for example, due to applied magnetic field. 
% As we briefly discussed in the  previous section, 
%the leading order perturbation from the magnetic field is a three-spin interaction  $\kappa$ which does not excite $Z_2$ flux but 
%opens a gap in the Majorana fermion bands with Chern number $\pm 1$~\cite{Kitaev2006}. As the gap at the Dirac points are $|\Delta_{\pm K}|\sim h^3/J_K^2$, when $|\Delta_{\pm K}|\lesssim v_F q/2$, the sound attenuation coefficient is not changed qualitatively. On the other hand, the time reversal symmetry breaking induces  the 
% mixing of the transverse and longitudinal phonon modes through the off-diagonal component of the polarization bubble, which leads to the non-zero  Hall viscosity  and the renormalization of the phonon velocity.

\subsection{Hall viscosity coefficient with time-reversal symmetry breaking}
\label{Appsec:Hall}
 We start by deriving the Hall viscosity coefficient $\eta_H$  which we introduced in Sec.~\ref{EltheoryTRSbroken}. We first recall %Eq. \ref{eq:Spha} 
 that it comes from the imaginary part of the off-diagonal component of the phonon polarization bubble, which is antisymmetric in exchanging the polarization indices $\mu\nu$. This process is non-dissipative as only the symmetric part of the viscosity coefficients contributes to dissipation~\cite{LandauElasticity,Avron1995,Read2011,Barkeshli2012}. Consequently, $\im \Pi_{\mathrm{ph}}(\qv,\Omega)$ does not necessarily come from the pole in $\Pi_{\mathrm{ph}}(\qv,\Omega)$, which involves on-shell scattering and is generally dissipative.
 
In the following, we consider the contribution when on-shell processes are not involved, which is called the ``intrinsic contribution"  in the study of the anomalous Hall effect~\cite{NagaosaAHErev}. It turns out that the intrinsic contribution and the scattering contribution can be separated when $v_s<v_F$. Whereas the intrinsic contribution is non-zero already at $T=0$, the scattering contribution requires $T>0$ for the reason we discussed in Sec.~\ref{attenT}. We also note that as there is no kinetimatic constraint for the intrinsic contribution, both the low-energy and high-energy Majorana fermions contribute to the polarization bubble. 
 % {\cred As we show in the following, the imaginary factor $i$ comes from the numerator due to time-reversal symmetry breaking.}
%Compared to Appendix~\ref{Appsec:Attenuation} in finding $\alpha_s$, there are two major differences in the computation. \emph{First}, a finite temperature is not necessary to obtain non-zero Hall viscosity, and in the following, we calculate $\eta_H$ at $T=0$. \emph{Second}, both the low energy and high energy Majorana fermions contribute to the polarization bubble. 
%As we show in the following, the imaginary factor $i$ comes from the numerator due to time-reversal symmetry breaking.}
% reference on the dissipation:
% Read 1008.0210v2 Read2011
% Avron https://arxiv.org/pdf/physics/9712050.pdf
% see also Landau v10 for a generic discussion
% intrinsic: Nagaosa NagaosaAHErev

In the following, we restrict our analysis to $T=0$. From Eqs.~\eqref{Appeq:C10} and \eqref{MFGF}, we note that at $T=0$ only the convolution of fermion propagators with $s s'<0$ are nonzero.   Similarly to how the sound attenuation coefficient was obtained in the preceding section, it is convenient to   integrate over frequencies first. We obtain:
\begin{align}
\mathbb{P}^{\pm,\mp}(\kv,\qv,i\Omega)&=\int\frac{\diff \omega}{2\pi} \frac{1}{i\omega\pm |\nf{\kv}|}\frac{1}{i(\omega+\Omega)\mp |\nf{\kv-\qv}|}\non\\&=\frac{1}{i\Omega\mp (|\nf{\kv-\qv}|+|\nf{\kv}|)}.
\label{eq:E1}
\end{align}
%where the upper indices $\pm,\mp$ indicate the combinations  of signs corresponding to $s s'<0$.
%The MFPh coupling follows from Eqs.~\eqref{eq:Vertices} and~\eqref{eq:lambdas}. Similar to how the sound attenuation coefficient was obtained, to compute the polarization bubble, it is convenient to first compute 
%\new{As was discussed in the main text, the MFPh coupling vertices originated from the time-reversal symmetry breaking spin Hamiltonian is smaller than Eqs.~\eqref{eq:Vertices} and~\eqref{eq:lambdas} by a factor $\kappa/J_k\sim(h/J_K)^3$, so we ignore in the following calculations.}

%Next, we compute the off-diagonal component of the polarization bubble by substituting  the MFPh coupling vertices with Pauli matrices  in sublattice space and get
Next, we compute the off-diagonal component of the polarization bubble. The MFPh coupling follows from Eqs.~\eqref{eq:Vertices} and~\eqref{eq:lambdas}. %Similar to how the sound attenuation coefficient was obtained, 
To compute the polarization bubble, it is again convenient to first compute
\begin{align}
\tilde \Pi^{\alpha\beta}(\qv,\kv,i\Omega)&=\tr{[\tau^{\alpha}\mathcal{G}(\kv,i\omega)\tau^{\beta}\mathcal{G}(\kv-\qv,i(\omega+\Omega))]},
\end{align}
where $\tau^{\alpha,\beta}$ are Pauli matrices in sublattice space, and $\tr[...]$ sums over Matsubara frequency and sublattice degrees of freedom. We have
\begin{widetext}
\begin{align}
\tilde \Pi_{\alpha\beta}(\kv,\qv,i\Omega)=& \frac{-i\Omega}{\Omega^2+(\df{\kv}+\df{\kv-\qv})^2}\left\{ -\tr{\left[\tau^\alpha\tau^\beta \frac{\nf{\kv-\qv}\cdot \vec{\tau}}{2\df{\kv-\qv}}\right]} +\tr{\left[\tau^\alpha \frac{\nf{\kv}\cdot \vec{\tau}}{2\df{\kv}}\tau^\beta\right]} \right\} \non\\
 & \frac{- (\df{\kv}+\df{\kv-\qv})}{\Omega^2+(\df{\kv}+\df{\kv-\qv})^2}\left\{ \frac{1}{2}\tr{\left[\tau^\alpha\tau^\beta\right]} -\frac{1}{2}\tr{\left[\tau^\alpha \frac{\nf{\kv}\cdot \vec{\tau}}{2\df{\kv}}\tau^\beta\frac{\nf{\kv-\qv}\cdot \vec{\tau}}{2\df{\kv-\qv}}\right]} \right\} .
 \label{Appeq:E3}
\end{align}
\end{widetext}
The first line (hereafter referred to as contribution I) and  the second line (contribution II) have different physical meanings. Contribution I contains a term linear in $\Omega$, and is odd under time-reversal, i.e.\ $\Omega\rightarrow -\Omega$ (in terms of both Matsubara frequency and time-ordered response in real frequency). Physically, this means that I contributes to the low-energy effective phonon action when time-reversal symmetry is broken.  Contribution II renormalizes the real part of the phonon propagator, e.g.\ the sound velocity, which  is a small effect and thus we do not consider it further.

The above observations are manifest by computing $\Pi_{\mathrm{ph}}^{\perp\parallel}(\qv,i\Omega)$ using the vertex functions $\hat \lambda^{\parallel(\perp)}_{{\bf q},{\bf k}}$ and summing ${\bf k}$ over the whole Brillouin zone. The MFPh coupling vertices originating from the time-reversal symmetry breaking spin Hamiltonian are ignored because they are smaller than those in Eq.~\eqref{eq:lambdas} by a factor of $\kappa/J_K\sim(h/J_K)^3$.
 Computing the contribution to $\Pi_{\mathrm{ph}}^{\perp\parallel}(\qv,i\Omega)$ from part (I) in Eq.~\eqref{Appeq:E3}, which leads to the Hall viscosity term, we get
\begin{widetext}
\begin{align}
\Pi_{\mathrm{ph}}^{\perp\parallel}(\qv,i\Omega)&=-\tr{[\hat\lambda^{\parallel}_{ \qv, \kv}\mathcal{G}(\kv,i\omega)\hat\lambda^{\perp}_{ \qv, \kv}\mathcal{G}(\kv-\qv,i(\omega+\Omega))]}\non\\
&= -q^2 \tr{ [
(\lambda^{\parallel}_{x,\kv,\qv}\tau^x +\lambda^{\parallel}_{y,\kv,\qv}\tau^y)\mathcal{G}(\kv,i\omega)(\lambda^{\perp}_{x,\kv,\qv}\tau^x +\lambda^{\perp}_{y,\kv,\qv}\tau^y)\mathcal{G}(\kv-\qv,i(\omega+\Omega))
]}\non\\
%&=-q^2\{
%\tr{[\lambda^{\parallel}_{x,\kv,\qv} \lambda^{\perp}_{y,\kv,\qv}\tau^x \mathcal{G}(\kv,i\omega) \tau^y \mathcal{G}(\kv-\qv,i(\omega+\Omega))]}+
%\tr{[\lambda^{\parallel}_{y,\kv,\qv} \lambda^{\perp}_{x,\kv,\qv}\tau^y \mathcal{G}(\kv,i\omega) \tau^x \mathcal{G}(\kv-\qv,i(\omega+\Omega))]}
%\}\non\\
&=2q^2 \Omega\int \frac{\diff^2 k}{\mathcal{A}_{\text{BZ}}} \frac{1}{\Omega^2+4\df{\kv}^2}\frac{\Delta_\kv}{\df{\kv}} (\lambda^{\parallel}_{x,\kv,\qv} \lambda^{\perp}_{y,\kv,\qv}-\lambda^{\parallel}_{y,\kv,\qv} \lambda^{\perp}_{x,\kv,\qv})
\label{eq:Polarization}
\end{align}
\end{widetext}
where we have used $\tr{[\tau^\alpha \tau^\beta \tau^\gamma]}=2i\epsilon^{\alpha\beta\gamma}$, and defined $\lambda^{\parallel(\perp)}_{x,y}$ as the $\tau^{x,y}$ components of the MFPh coupling vertices $\hat \lambda^{\parallel(\perp)}_{{\bf q},{\bf k}}$ :

\begin{align}
\label{eq:lambdaxy}
\hat \lambda^{\parallel}_{x,{\bf q},{\bf k}}=&\frac{i\lambda_{A_{1}}}{2} q 
f''_{\kv}+\frac{i\lambda_{E_{2}}}{2} q \left(
 \cos 2\theta_q f''_{1,\bf k}
 +\sin 2\theta_q f''_{2,\bf k}
\right),\non\\
\hat \lambda^{\parallel}_{y,{\bf q},{\bf k}}=&\frac{i\lambda_{A_{1}}}{2} q 
f'_{\kv}+\frac{i\lambda_{E_{2}}}{2} q \left(
 \cos 2\theta_q f'_{1,\bf k}
 +\sin 2\theta_q f'_{2,\bf k}
\right),\non\\
\hat \lambda^{\perp}_{x,{\bf q},{\bf k}}=&\frac{i\lambda_{E_{2}}}{2} q \left(
 -\sin 2\theta_q f''_{1,\bf k}
 +\cos 2\theta_q f''_{2,\bf k}
\right),\non\\
\hat \lambda^{\perp}_{y,{\bf q},{\bf k}}=&\frac{i\lambda_{E_{2}}}{2} q \left(
 -\sin 2\theta_q f'_{1,\bf k}
 +\cos 2\theta_q f'_{2,\bf k}
\right).
\end{align} 
Note that from Eq.~(\ref{eq:Polarization})  one  clearly sees that the contribution to $\Pi_{\mathrm{ph}}^{\mu\nu}(\qv,i\Omega)$ from the  part (I) vanishes for $\mu=\nu$, and  that $\Pi_{\mathrm{ph}}^{\parallel\perp}=-\Pi_{\mathrm{ph}}^{\perp\parallel}$. Furthermore, since $\Delta_\kv \propto \kappa $, the contribution from Eq.~(\ref{eq:Polarization}) only exists when time-reversal symmetry is broken (i.e. when $\kappa \neq 0$).

The main contribution to the momentum integration is near the Dirac points ($\pm K$). Linearizing the spectrum near the Dirac points gives $\df{\pm K +\kv}=\{\mp 3 J_K k_y,-3J_K k_x,6\sqrt{3}\kappa\}$, and Eq.~(\ref{eq:Polarization}) becomes
\begin{align}
\Pi_{\mathrm{ph}}^{\perp\parallel}(\qv,i\Omega)\approx-q^2 \Omega \frac{\lambda_{E_2}^2}{J_K^2}\frac{3\sqrt{3}}{4\pi} \sgn{\kappa}.
\label{eq:mixingM}
\end{align}
We note that the result is ``quantized" in the sense that it does not depend on microscopic details of the time-reversal symmetry breaking perturbation, i.e.\ the value of $\kappa$, but only on its sign.  
In fact, one can show that it is essentially the same setting to compute Hall conductivity of the anomalous quantum Hall effect when the MFPh vertex is constrained to near $\pm K$~\cite{Rosch2018}. 
%Eq.~(\ref{eq:mixingM}) recovers the results  obtained in Ref.~\cite{Rosch2018} where
%  the Hall conductivity of anomalous quantum Hall effect was computed with the MFPh vertices  evaluated near  the Dirac points $\pm K$.
On the other hand, the analogy breaks down beyond the linear approximation of the Majorana fermion spectrum,   since the high energy Majorana fermions also contribute to the Hall viscosity. After integrating over the whole Brillouin zone, we find that the numerical prefactor depends on the magnitude of $\kappa$, i.e., on the magnitude of the Dirac mass.  

\begin{figure}[t]
\centering
\includegraphics[width=0.9\columnwidth]{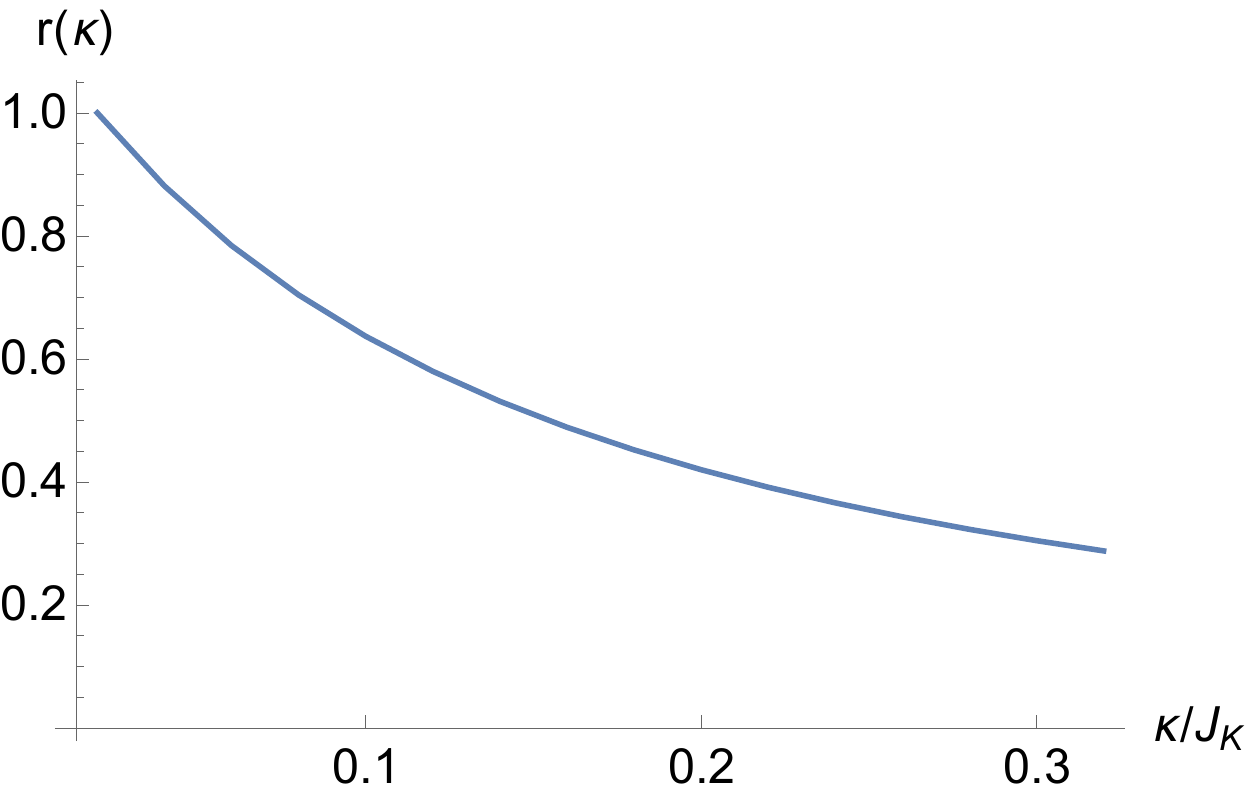}
\caption{Plot of $r(\kappa)$ as a function of the dimensionless ratio $\kappa/J_K$}.\label{fig:rk}
\end{figure}

In the following, we define the result found in Eq.~\eqref{eq:mixingM} as $\Pi_{\mathrm{ph}}^{\perp\parallel}(q,i\Omega)_{\kappa\rightarrow 0}$, since in
the limit $|\kappa|/J_K\rightarrow 0^+$, Eq.~\eqref{eq:mixingM} is recovered. This is because the details of the Majorana fermion spectrum are not relevant in this limit, so the result should be the same as that obtained from the approximation of linearized spectrum. 
 The polarization bubble at  finite $\kappa$ can be expressed as $\Pi_{\mathrm{ph}}^{\perp\parallel}(q,i\Omega)_\kappa=r(\kappa)\,\Pi_{\mathrm{ph}}^{\perp\parallel}(q,i\Omega)_{\kappa\rightarrow 0}$, where  the numerically evaluated  coefficient $r(\kappa)$ is shown in Fig.~\ref{fig:rk}. We note from the numerical calculation that only the $\lambda_{E_2}^2$ term in Eq.~\eqref{eq:Polarization} contributes to the Hall viscosity coefficient.

The time-ordered phonon polarization bubble in real frequency can be obtained by analytic continuation. We have  
\begin{align}
\im \Pi_{\mathrm{ph}}^{\perp\parallel}(\qv,\Omega)_\kappa = q^2 \Omega \frac{\lambda_{E_2}^2}{J_K^2}\sgn{\kappa}\frac{3\sqrt{3}}{4\pi} r(\kappa).
\label{eq:PolarizationH}
\end{align}

 Here we need to stress that while both Eq.~\eqref{eq:PolarizationH}, leading to the Hall viscosity, and the imaginary part of the \emph{real-time-ordered} phonon polarization bubble associated with the sound attenuation as we found in Eqs.~\eqref{Piparallel} and~\eqref{Piperp}  are linear in $\Omega$, they are fundamentally different, i.e., the former is non-dissipative and the latter, dissipative. As was emphasized in our microscopic derivation, this difference comes from  the distinct origins of the microscopic processes contributing to  the imaginary part: One comes from the poles of the polarization bubble and the other from the numerator $\sim \tr{[\tau^x\tau^y\tau^z]}$. From a phenomenological perspective, the difference comes from the fact that the imaginary part of the time ordered polarization bubble associated with dissipation is even under $\Omega\rightarrow -\Omega $ due to causality, i.e.\ $\sim |\Omega|$.  
 %{\cred The non-analytic feature is associated with the Landau damping of the boson, see e.g.~\cite{Abanov2003}.} 
 On the other hand, the imaginary part of the time ordered polarization bubble associated with the Hall viscosity is odd under $\Omega\rightarrow -\Omega$, i.e.\ $\sim \Omega$. This is consistent with the causality constraint because the Hall viscosity term only appears in the off-diagonal component of the phonon polarization $\Pi_{\mathrm{ph} }^{\mu\nu} (\qv, \Omega)$.

The Hall viscosity coefficient $\eta_H$ in Eq.~\eqref{etaH} can now  be related to the phonon polarization bubble by expressing Eq.~\eqref{Appeq:C10} in terms of
\begin{align}
\Pi_{\mathrm{ph}}^{\mu\nu}(\qv,\Omega)=\frac{-i}{2!}\int \diff t\, e^{i\Omega t} \langle T\,  \frac{\partial H_c}{\partial \tilde u_{\qv,\mu}}(t) \, \frac{\partial H_c}{\partial \tilde u_{-\qv, \nu}} (0) \rangle,
\end{align}
where we recall  that $H_c$ is the spin-lattice coupling Hamiltonian. Comparing with the expression in Eq.~\eqref{etaH}, we find the Hall viscosity coefficient  to be equal to
\begin{align}
\eta_H=-\frac{1}{\ell_a^2}\frac{\lambda_{E_2}^2}{J_K^2}\sgn{\kappa}\frac{3\sqrt{3}}{2} r(\kappa).
\end{align}

Finally, we also note that the Hall viscosity term may be the starting point to obtain the phonon thermal Hall conductivity, which is finite and of order $\sim T^3$~\cite{Shi2012,Rosch2018}.

\subsection{Renormalization of the phonon spectrum}

Next we  discuss the renormalization  of the phonon spectrum due to $\Pi_{\mathrm{ph}}^{\parallel\perp}(\qv,\Omega)$ that mixes  the longitudinal and transverse phonon modes. 

For brevity in the expression, we write Eq.~\eqref{eq:PolarizationH} as $\im \Pi_{\mathrm{ph}}^{\perp\parallel}(\qv,\Omega)_\kappa = q^2 \Omega\,g(\kappa)$, where
\begin{align}
 g(\kappa)\equiv\frac{\lambda_{E_2}^2}{J_K^2}\frac{3\sqrt{3}}{4\pi} \sgn{\kappa}\,r(\kappa). 
 \end{align}
 The renormalized phonon propagator can be expressed in matrix form as
\begin{align}
 D^{-1}(\qv,\Omega)=-\rho\delta_V 
\begin{pmatrix}
\Omega^2-(v_s^{\parallel}q)^2 & i q^2\Omega g(\kappa)/(\rho\delta_V)\\
-i q^2\Omega g(\kappa)/(\rho\delta_V) & \Omega^2-(v_s^{\perp}q)^2
\end{pmatrix}.
\label{Phpropagator}
\end{align}
To the leading order in $g(\kappa)$, the renormalized spectrum is given by
\begin{align}\label{eq:ren-spectrum}
\Omega_{\mu,\qv}&=v_s^{\mu}q\sqrt{1+q^2 \frac{(g(\kappa)/\rho\delta_V)^2}{(v_s^{\mu})^2-(v_s^{\nu})^2}}\non\\&\sim 
v_s^{\mu}q \left[1+\frac{1}{2}q^2 \frac{(g(\kappa)/\rho\delta_V)^2}{(v_s^{\mu})^2-(v_s^{\nu})^2} \right].
\end{align}
Note that the correction from the $q^3$ term changes sign for the longitudinal and transverse phonon modes, so the spectrum bends upwards(downwards) for the longitudinal(transverse) mode [see Fig. \ref{fig:summary} (c)].  To determine the energy scale (relative to the Kitaev interaction $J_K$) when the deviation from the linear spectrum becomes prominent, the second term inside the square brackets on the right-hand side of Eq.~(\ref{eq:ren-spectrum}) can be written as
\begin{align}
&\frac{1}{2}q^2 \frac{(g(\kappa)/\rho\delta_V)^2}{(v_s^{\mu})^2-(v_s^{\nu})^2}\non\\
&=\frac{1}{2}\left(\frac{v_s^{\mu}q}{J_K}\right)^2\left(\frac{\lambda_{E_2}^2}{J_K C_{\mu}\delta_V}\right)^2 \left(\frac{(v_s^{\mu})^2}{(v_s^{\mu})^2-(v_s^{\nu})^2}\right) \left(\frac{3\sqrt{3}}{4\pi} r(\kappa)\right)^2,
\label{appeq:Hallcubic}
\end{align}
where the relation between the sound velocity, the mass density, and the components of the elastic modulus tensor was defined in Eq.~\eqref{appeq:PhSpectrum}, and $C_{\parallel}=C_1+C_2$, and $C_{\perp}=C_2$. It is straightforward to see that each terms in large parentheses in Eq.~\eqref{appeq:Hallcubic} represents a dimensionless quantity. Thus, when Eq.~\eqref{appeq:Hallcubic} is comparable to 1, the cubic term in $q$ becomes important. Therefore, when $v_s^{\perp}<v_s^{\parallel}$, the mixing shifts the transverse mode spectrum downwards and the longitudinal mode spectrum upwards at a characteristic scale
\begin{align}
\label{eq:mixingc}
\delta_c&=
\left(\frac{v_s^{\parallel}q}{J_K}\right)_c \sim \frac{J_K C_\parallel \delta_V}{\lambda_{E_2}^2}
\frac{\sqrt{(v_s^\parallel)^2-(v_s^{\perp})^2}}{v_s^\parallel}\non\\& =
\frac{J_K \sqrt{C_\parallel(C_\parallel-C_\perp)} \delta_V}{\lambda_{E_2}^2}.
\end{align} 
We note that in general $C\delta_V$ is a large energy scale of order eV, so large spin-lattice coupling is necessary to see appreciable change of the phonon spectrum due to the time-reversal breaking $\kappa$-term. However, more knowledges of the energy scales associated with $\lambda_{E_2}$ and $\sqrt{C_\parallel(C_\parallel-C_\perp)}\delta_V$ in Kitaev materials is needed to have a quantitative estimate of $\delta_c$.

%as the linear in $\Omega$ term in a dissipative bosonic system requires analyticitity, i.e. $\im \Pi \sim |\Omega|$. 
%
%\begin{align}
%\tilde \Pi^{\alpha\beta}(q,i\Omega)&=\tr{[\tau^{\alpha}\mathcal{G}(k,i\omega_m)\tau^{\beta}\mathcal{G}(k-q,i(\omega_m+\Omega_n))]}
%\end{align}
%where $\alpha,\beta=x,y$. 
\section{Summary and discussion}\label{Sec:summary}
In this work, we studied the effects of the spin-lattice coupling on the acoustic phonon dynamics in the Kitaev  spin liquid and obtained various observables, such as  the sound attenuation ($\alpha_s$), the Hall viscosity coefficient ($\eta_H$)  and the renormalized 2D acoustic phonon spectrum [see Eq. (\ref{eq:ren-spectrum})]. We suggest that if  measured,  these  observable effects can be used as potential probes of spin fractionalization in the Kitaev materials.

We demonstrated that these observables can be obtained from the phonon polarization bubble. To compute the bubble, a microscopic
 low-energy effective spin-lattice coupling was derived and a diagrammatic computation procedure was formulated in terms of the matter Majorana fermions and the acoustic phonons. 
We first showed that the sound attenuation comes from the phonon scattering off  the Majorana fermions. Due to the temperature dependence of the scattering phase space, $\alpha_s\propto T$, which distinguishes from the attenuation in other interaction channels, such as proportional to $T^3$ due to anharmonic phonon interactions. We also showed that $\alpha_s$ is highly anisotropic with an angular modulation proportional to $(1\pm \cos 6\theta_q)$, which should be attributed to the combined effects of the spin-lattice coupling and the low-energy Dirac spectrum of the Majorana fermions. We then computed  the Hall viscosity coefficient of the phonon effective action, which is non-zero when time-reversal symmetry is broken. 
%In particular, due to the spin-lattice coupling, the time-reversal breaking spin Hamiltonian can induce the phonon Hall viscosity.
Different from electron Hall fluids, the Hall viscosity coefficient is not quantized. We found two contributions: one that is non-perturbative and resembles the quantized Hall conductance, and another that is perturbative in the time-reversal breaking term $\kappa$. The Hall viscosity coefficient may be probed indirectly from the phonon spectrum, where it is manifested as a deviation of the phonon dispersion from linear to non-linear near a characteristic momentum $q_c$, which depends on the strength of the spin-lattice coupling [See Eq.~\eqref{eq:mixingc}]. Of course, such a deviation is generically expected for large enough momenta, where our long wavelength approximation breaks down. Our key point is that this new scale $q_c$ where the deviation from linear to non-linear takes place, is only present in the presence of an applied magnetic field. i.e. when time-reversal symmetry is broken. To obtain a quantitative estimate of the sound attenuation coefficient $\alpha_s$ and characteristic momentum $q_c$, we call for studies of the spin-lattice coupling and elastic modulus tensor of Kitaev materials from experiments and \textit{ab initio} studies. 

%we expect that some qualitative features remain valid up to a higher temperature cutoff than the one set by the flux gap. There are two main reasons for this. First, the linear in T scaling of the sound attenuation coefficient comes from the general phase space analysis. For dilute flux excitation configurations, as the Majorana fermion spectrum is not modified much, the linear in T behavior qualitatively holds. Second, our analysis shows a clear angular anisotropy of the sound attenuation coefficient, and the Dirac spectrum is partly responsible for this anisotropy. Thermal flux excitations may smear the anisotropy, but it should be present up to a higher temperature. To better justify the motivation, we added a sentence at the end of the main text

As a final remark, we would like to emphasize that only in the temperature range $T<\Delta_{\mathrm{flux}}$, which we focused in this work, can the thermal flux excitations be ignored. On the other hand, the key qualitative features we found, such as the linear in $T$ dependence and angular anisotropy of the sound attenuation coefficient, should remain valid up to a higher temperature cutoff than the one set by the flux gap. Understanding the effects of spin-lattice coupling in the temperature range beyond the constraint set by $\Delta_{\mathrm{flux}}$ would be highly desirable~\cite{Brenig2019}, and we leave it for future work. 

% the model, effective action, calculation part: vertex, phonon propagator, attenuation, Hall viscosity term
% discussion part: without flux excitations, 
%%%%%%%%%%%%%%%%%%%%%%%%%%%%%%%%%%%%%%%%%%%%%%%%%%%%%%%%%

%\vspace*{0.3cm} 
\section*{Acknowledgments} 
We thank  Sananda Biswas, Wolfram Brenig, Fiona Burnell, Itamar Kimchi, Sai Mu, Lucile Savary, Xuzhe Ying, Roser Valenti and Sergei Zherlitsyn for helpful discussions. M.Y.\ is grateful for support from the University of Minnesota through a Louise Dosdall Fellowship, the Army Research Office MURI Grant No.\ ARO W911NF-16-1-0361, (Floquet engineering and metastable states), and the National Science Foundation under Grant No.\ NSF PHY-1748958 from UC Santa Barbara. R.M.F.\ was supported by the U.S.\ Department of Energy, Office of Science, Basic Energy Sciences, Materials Sciences and Engineering Division, under Award No.\ DE-SC0020045. The work of  N.B.P.\ was supported by the U.S.\ Department of Energy, Office of Science, Basic Energy Sciences under Award No.\ DE-SC0018056. N.B.P.\ also acknowledges the hospitality  of the KITP and NSF Grant No.\ PHY-1748958.

\appendix
\onecolumngrid
%%%%%%%%%%%%%%%%%%%%%%%%%%%%%%%%%%%%%%%%%%%%%%%%%%%%%%%%%
%%%%%%%%%%%%%%%%%%%%%%%%%%%%%%%%%%%%%%%%%%%%%%%%%%%%%%%%%
%%%%%%%%%%%%%%%%%%%%%%%%%%%%%%%%%%%%%%%%%%%%%%%%%%%%%%%%%

\section{Majorana fermion propagator}\label{MajoranaGF}
%\subsection{Free Majorana fermion Green's function}
In this section, we briefly review the procedure of obtaining the Majorana fermion Green's function, i.e.\ Eq.\eqref{MFGF} in the main text.
It can be done  either by the diagonalization of  the Hamiltonian Eq.(\ref{eq:Kmodel1})
 directly keeping track of the independent Majorana  fermion modes or by expressing   Majorana fermions  in terms of complex fermions. In the following, we take the second approach and  
% It has also been checked that the two ways give consistent results.
% and diagonalize the complex fermions to avoid the possible confusion in mode counting. 
express $\{c_{\ve{r},A},c_{\ve{r},B}\}$ in terms of complex fermions $\gamma_\ve{r}$ on the bond as
$c_{\ve{r},A}=(\gamma_\ve{r}+\gamma^{\dg}_\ve{r})$ and $ c_{\ve{r},B}=i(\gamma_\ve{r}-\gamma^{\dg}_\ve{r})$.
Performing the Fourier transformation to the momentum space,  $\gamma_{\bf k}=\frac{1}{\sqrt{N}}\sum \gamma_\ve{r}e^{-i{\bf k}\cdot {\ve r}}$,
we can  relate $\gamma_{\bf k},\gamma^{\dg}_{\bf k}$  with $a_{\bf k},b_{\bf k}$  as
$
\gamma_{\bf k}=\frac{1}{\sqrt{2}}(a_{\bf k}-i b_{\bf k})$ and $ \gamma^{\dg}_{\bf k}=\frac{1}{\sqrt{2}}(a_{-{\bf k}}+i b_{-{\bf k}})$
%The Kitaev model Eq.(\ref{eq:Kmodel1}) in the zero-flux sector expressed in terms of $\gamma$ fermions is  now given by
and thus obtain
\begin{eqnarray}  
\tilde H_s=\frac{1}{2}\sum_{\bf k} 
\psi^T_{-\ve{k}}
(-\tau_x\im f_{\bf k}-\tau_y\re f_{\bf k} +\tau_z \Delta_{\bf k})
\psi_{\ve k}\rightarrow
 \frac{1}{2}\sum_{\ve{k}}
\begin{pmatrix}
\gamma^{\dg}_{\bf k}, \gamma_{-{\bf k}}
\end{pmatrix}
\begin{pmatrix}
-f'_{\bf k} & i f''_{\bf k} +\Delta_{\bf k} \\
-i f''_{\bf k}k+\Delta_{\bf k}  & f'_{\bf k}
\end{pmatrix}
\left(\begin{array}{c}
\gamma_{\bf k} \\ \gamma^{\dg}_{-{\bf k}}\end{array}\right),
\label{eq:Kmodelgamma}
\end{eqnarray}
which can be diagonalized through Bogoliubov transformation. The Hamiltonian in terms of Bogoliubov fermions $\beta_{\bf k}$ is
\begin{align}
\tilde H_s=\frac{1}{2}\sum_{\ve{k}}
\begin{pmatrix}
\beta^{\dg}_{\bf k}, \beta_{-{\bf k}}
\end{pmatrix}
\begin{pmatrix}
\df{{\bf k}} & 0\\
0 & -\df{{\bf k}}
\end{pmatrix}
\left(\begin{array}{c}
\beta_{\bf k} \\ \beta^{\dg}_{-{\bf k}}\end{array}\right),
\end{align}
from which we find $-i\langle T \beta^{\dg}_{\bf k}\beta_{\bf k}\rangle_\omega=\frac{1}{\omega+\df{{\bf k}}-i\delta}$ and $-i\langle T \beta_k \beta^{\dg}_{\bf k}\rangle_\omega=\frac{1}{\omega-\df{{\bf k}}+i\delta}$. Using  the transformation between $\psi_{\bf k}$ and $\beta_{\bf k}$, we obtain Eq.~(\ref{MFGF}).
%\subsection{Majorana Fermion self-energy} \label{Majoranaselfenergy}
%The Majorana fermion self-energy [Fig.\ref{fig:Attenuation} (c) of the main text] can be expressed as
%\begin{align}
%\Sigma({\bf k},\omega)&=i\tr{[\hat\lambda^{\nu}_{ {\bf q}, {\bf k}}D_{\nu\nu}^{(0)}({\bf q},\Omega)\hat\lambda^{\nu}_{ {\bf q}, {\bf k}}\mathcal{G}({\bf k}-{\bf q},\omega-\Omega)]},
%\label{Appeq:MFSE}
%\end{align}
%where the bare  phonon propagator in terms of the lattice displacement field $\tilde{u}_{\ve{q},\nu}$ is given by  Eq.(\ref{BarePHpropagator}).
% In the MFPh coupling vertex, we only keep terms linear in ${\ve q}$ and ${\ve k}$, so  
% $\hat\lambda_{{\bf q},{\bf k}-{\bf q}}\simeq \hat\lambda_{{\bf q},{\bf k}}$. 
% $\tr[...]$ sums over momentum ${\ve q}$, polarization $\nu$ and integrates over frequency as $\int \frac{\diff \Omega}{2\pi}$.

%%%%%%%%%%%%%%%%%%%%%%%%%%%%%%%%%%%%%%%%%%%%%%%%%%%%%%%%%
\section{Relationship between the sound attenuation coefficient $\alpha_s$ and the phonon polarization buble}
\label{Appsec:alphas}

To derive Eq. (\ref{eq:Attenuation}), we consider first the attenuation coefficient of a 1d system in which the sound wave propagates in a given direction. We start with the lossy sound wave equation~\cite{Plee2011}
\begin{align}
v_s^2(1+\tau_s \frac{\partial}{\partial t})\nabla^2 \ve{u} = \frac{\partial^2}{\partial t^2} \ve{u},
\label{SMeq:1-1}
\end{align}
where $\tau_s$ is the sound wave relaxation time.
With a plane wave ansatz for $\ve{u}(x,t)=\ve{u}_0 e^{i(\Omega t - q \, x)}$, we find that
\begin{align}
q=\frac{\Omega}{v_s}\frac{1}{\sqrt{1+i\Omega \tau_s}}\approx \frac{\Omega}{v_s}(1-\frac{i \Omega \tau_s}{2}).
\label{SMeq:1-1b}
\end{align}
 Then the  lossy sound wave  becomes
 \begin{align}
\ve{u}(x,t)\approx e^{i (\Omega t - \frac{\Omega}{v_s} x)}e^{-\frac{\Omega^2\tau_s}{2 v_s} x}
\end{align}
  and  the sound attenuation coefficient can be simply written as
\begin{align}
\alpha_s=\frac{\Omega^2\tau_s}{2 v_s}.
\label{SMeq:1-2}
\end{align}
Following Eq.~\eqref{SMeq:1-1}, the phonon propagator can be expressed as
\begin{align}
D^{-1}(q,\Omega)= \rho\delta_V(\Omega^2-v_s^2 q^2-i\Omega \tau_s v_s^2 q^2).
\label{eq:PhononPole}
\end{align} 

On the other hand, the inverse of the  renormalized  phonon propagator  Eq.~\eqref{renphpropagator}
can be written as
\begin{align}
D^{-1}(q,\Omega)= \rho\delta_V(\Omega^2-v_s^2 q^2)+i\im \Pi_{\mathrm{ph}}(q,\Omega)
\label{SMeq:1-3}
\end{align} 
where the real part of  the phonon self-energy $\re \hat\Pi_{\mathrm{ph}}(q,\Omega)$  has been absorbed into the renormalization of sound velocity $v_s$.
Equating Eqs.~\eqref{SMeq:1-3} and~\eqref{eq:PhononPole} allows us to relate the life time $\tau_s$  to 
 the imaginary part of phonon self-energy as
\begin{align} 
 \tau_s=-\frac{1}{\Omega v_s^2 q^2 \rho\delta_V}\im \Pi_{\mathrm{ph}}(q,\Omega)
 \label{SMeq:1-4}
\end{align} 
  and through this relate the sound attenuation coefficient  with the imaginary part of phonon self-energy:
 % The last one in Eq.~\eqref{SMeq:1-3} comes from the relation of $\Omega$ and $k$ from Eq.~\eqref{SMeq:1-1}, which should match the pole of the phonon Green's function. From Eq.~\eqref{SMeq:1-2} and~\eqref{SMeq:1-3}, we find Eq.~(2) in the MS, 
%which we repeat below
\begin{align}\label{SMeq:1-5}
\alpha_s(q)=-\frac{1}{2 v_s \Omega_q \rho\delta_V}\im \Pi_{\mathrm{ph}}(q,\Omega_q),
\end{align}
where $\Omega_q=v_s q$.
We note that $\im \hat\Pi_{\mathrm{ph}}/(\rho\,\delta_V)$ is in unit of energy square, and $\alpha_s(q)$ is in unit of inverse distance. Eq.~\eqref{SMeq:1-5} can be straight-forwardly generalized to higher dimensions by replacing the scalar $q$ with $\qv$.\\

%%%%%%%%%%%%%%%%%%%%%%%%%%%%%%%%%%%%%%%%%%%%%%%%%%%%%%%%%

%%%%%%%%%%%%%%%%%%%%%%%%%%%%%%%%%%%%%%%%%%%%%%%%%%%%%%%%%
\section{Technical details of the evaluation of the sound attenuation coefficient $\alpha_s$}
\label{Appsec:Attenuation}

 As we discussed in  Sec.\ref{attenT},  when    $T< \Delta_{\mathrm{flux}}\ll J_K$ the spectrum of the Majorana fermions can be  linearized near  the Dirac points $\pm K$  and  the MFPh vertices $\hat \lambda^{\nu}_{{\bf q},{\bf k}}$  are constant in terms of $\ve k$. Thus,  the momentum summation in Eq.~\eqref{Appeq:Polarization} can be obtained by first  replacing the true MFPh vertices with the Pauli matrices $\tau^{x,y}$, i.e.\ Eq.~\eqref{eq:Polarizationxy},
\begin{align}
&\tilde \Pi_{\alpha\beta}(\qv,\Omega)
=\tr{[\tau^{\alpha}\mathcal{G}(\kv,i\omega_m)\tau^{\beta}\mathcal{G}(\kv-\qv,i(\omega_m+\Omega_n))]}_{i\Omega\rightarrow \Omega+i\delta\sgn\Omega}\non\\
=&\sum_{s,s'}\int \frac{\diff^2 k}{\mathcal{A}_{\text{BZ}}}\frac{1}{2}\frac{\big(s\tanh{\beta|\nf{\kv}|/2}-s'\tanh{\beta|\nf{\kv-\qv}|/2}\big)}{\Omega+s'|\nf{\kv-\qv}|-s|\nf{\kv}|+i\delta\sgn\Omega}\tr[\tau^{\alpha}(\frac{I}{2}-s\frac{\vec{\mathcal{F}}_\kv\cdot\vec{\tau}}{2|\nf{\kv}|})\tau^{\beta}(\frac{I}{2}-s'\frac{\vec{\mathcal{F}}_{\kv-\qv}\cdot\vec{\tau}}{2|\nf{\kv-\qv}|})]\non\\
=&\int \frac{\diff^2 k}{\mathcal{A}_{\text{BZ}}}\frac{1}{2}\big(\tanh{\frac{\beta|\nf{\kv}|}{2}}-\tanh{\frac{\beta|\nf{\kv-\qv}|}{2}}\big)\{\frac{1}{\Omega+|\nf{\kv-\qv}|-|\nf{\kv}|+i\delta\sgn\Omega}\tr[\tau^{\alpha}(\frac{I}{2}-\frac{\vec{\mathcal{F}}_\kv\cdot\vec{\tau}}{2|\nf{\kv}|})\tau^{\beta}(\frac{I}{2}-\frac{\vec{\mathcal{F}}_{\kv-\qv}\cdot\vec{\tau}}{2|\nf{\kv-\qv}|})]\non\\
&\quad\qquad\qquad \qquad\qquad\qquad\qquad\qquad\qquad\quad -\frac{1}{\Omega-|\nf{\kv-\qv}|+|\nf{\kv}|+i\delta\sgn\Omega}\tr[\tau^{\alpha}(\frac{I}{2}+\frac{\vec{\mathcal{F}}_\kv\cdot\vec{\tau}}{2|\nf{\kv}|})\tau^{\beta}(\frac{I}{2}+\frac{\vec{\mathcal{F}}_{\kv-\qv}\cdot\vec{\tau}}{2|\nf{\kv-\qv}|})]\},
\end{align}
where $\alpha,\beta=x,y$, $\mathcal{A}_{\text{BZ}}=8\pi^2/(3\sqrt{3})$ is the area of the honeycomb lattice Brillouin zone.    Expanding $\vec{\mathcal{F}}_\kv$   and  $\vec{\mathcal{F}}_{\kv-\qv}$ near the Dirac points $\pm K$, we find 
\begin{align}
\label{Appeq:Polarizationxy}
\tilde \Pi_{\alpha\beta}(\qv,\Omega)&=\int \frac{\diff^2 k}{\mathcal{A}_{\text{BZ}}}\frac{1}{2}\big(\tanh{\frac{\beta|\nf{\pm K+\kv}|}{2}}-\tanh{\frac{\beta|\nf{\pm K+\kv-\qv}|}{2}}\big)\frac{1}{\Omega+|\nf{\pm K+\kv-\qv}|-|\nf{\pm K+\kv}|+i\delta\sgn\Omega}\times\non\\
&\qquad\qquad \quad\{\frac{1}{2}\tr[\tau^\alpha\tau^\beta]+\tr[\tau^\alpha\frac{\vec{\mathcal{F}}_{\pm K+\kv}\cdot\vec{\tau}}{2|\nf{\pm K+\kv}|}\tau^\beta\frac{\vec{\mathcal{F}}_{\pm K+\kv-\qv}\cdot\vec{\tau}}{2|\nf{\pm K+\kv-\qv}|}]+\tr[\tau^\alpha\frac{\vec{\mathcal{F}}_{\pm K-\kv+\qv}\cdot\vec{\tau}}{2|\nf{\pm K-\kv+\qv}|}\tau^\beta\frac{\vec{\mathcal{F}}_{\pm K-\kv}\cdot\vec{\tau}}{2|\nf{\pm K-\kv}|}]\}.
\end{align}

Using the identities for Pauli matrices, $\tr[\tau^\alpha\tau^\beta]=2\delta^{\alpha\beta}$ and $\tr[\tau^\alpha\tau^\beta\tau^\gamma\tau^\delta]=2(\delta^{\alpha\beta}\delta^{\gamma\delta}-\delta^{\alpha\gamma}\delta^{\beta\delta}+\delta^{\alpha\delta}\delta^{\beta\gamma})$, we find that the second line for $\tilde\Pi^{\alpha\beta}$ can be simplified and we get  
\begin{align}
\tilde \Pi_{xx}(\qv,\Omega)&\sim\{1+\frac{\mathcal{F}^x_{\pm K+\kv}\mathcal{F}^x_{\pm K+\kv-\qv}-\mathcal{F}^y_{\pm K+\kv}\mathcal{F}^y_{\pm K+\kv-\qv}}{|\nf{\pm K+\kv}||\nf{\pm K+\kv-\qv}|}-\frac{\mathcal{F}^z_{\pm K+\kv}\mathcal{F}^z_{\pm K+\kv-\qv}}{|\nf{\pm K+\kv}||\nf{\pm K+\kv-\qv}|}\},\non\\
\tilde \Pi_{yy}(\qv,\Omega)&\sim\{1-\frac{\mathcal{F}^x_{\pm K+\kv}\mathcal{F}^x_{\pm K+\kv-\qv}-\mathcal{F}^y_{\pm K+\kv}\mathcal{F}^y_{\pm K+\kv-\qv}}{|\nf{\pm K+\kv}||\nf{\pm K+\kv-\qv}|}-\frac{\mathcal{F}^z_{\pm K+\kv}\mathcal{F}^z_{\pm K+\kv-\qv}}{|\nf{\pm K+\kv}||\nf{\pm K+\kv-\qv}|}\},\non\\
\tilde \Pi_{xy}(\qv,\Omega)&\sim\{\frac{\mathcal{F}^x_{\pm K+\kv}\mathcal{F}^y_{\pm K+\kv-\qv}+\mathcal{F}^y_{\pm K+\kv}\mathcal{F}^x_{\pm K+\kv-\qv}}{|\nf{\pm K+\kv}||\nf{\pm K+\kv-\qv}|}\}.
\end{align}
The diagonal components of  $\im \Pi_{\mathrm{ph}}^{\nu\nu}(\qv,\Omega)$ 
in terms of $\tilde \Pi_{\alpha\beta}$  are then given by
\begin{align}
\Pi^{\parallel\parallel}(\qv,\Omega)&=9\lambda_{E_2}^2q^2[\sin^2 2\theta_q\tilde \Pi_{xx}+\cos^2 2\theta_q\tilde \Pi_{yy}-\sin 2\theta_q \cos 2\theta_q(\tilde \Pi_{xy}+\tilde \Pi_{yx})],\label{Piparallel-ap}\\
\Pi^{\perp\perp}(\qv,\Omega)&=9\lambda_{E_2}^2q^2[\cos^2 2\theta_q\tilde \Pi_{xx}+\sin^2 2\theta_q\tilde \Pi_{yy}+\sin 2\theta_q \cos 2\theta_q(\tilde \Pi_{xy}+\tilde \Pi_{yx})].\label{Piperp-ap}
\end{align}
 These are formula Eqs.~\eqref{PolarizationL} and~\eqref{PolarizationT} of the main text.

We first consider $\mathcal{F}^z=\Delta_\kv=0$, which is the case when $\kappa=0$. Linearizing  $\vec{\mathcal{F}}_{\pm K+\kv}$ near $K$ (or $-K$) as
\begin{align}
\vec{\mathcal{F}}_{K+\kv}=|f_{K+\kv}|\{-\sin\phi_k,-\cos\phi_k,0\}=v_F k \{-\sin\phi_k,-\cos\phi_k,0\},
\end{align}
 we find
\begin{align}
i\im \Pi_{\mathrm{ph}}^{\parallel\parallel}(\qv,\Omega)=&-i9\lambda_{E_2}^2q^2\pi\int \frac{\diff^2 k}{\mathcal{A}_{\text{BZ}}}\sgn \Omega\, \delta(\Omega+|f_{K+\kv-\qv}|-|f_{ K+\kv}|)\big(\tanh{\frac{\beta|f_{ K+\kv}|}{2}}-\tanh{\frac{\beta|f_{ K+\kv-\qv}|}{2}}\big)\times\non\\
&\quad(1-\cos 4\theta_q \frac{\mathcal{F}^x_{ K+\kv}\mathcal{F}^x_{ K+\kv-\qv}-\mathcal{F}^y_{ K+\kv}\mathcal{F}^y_{ K+\kv-\qv}}{|f_{ K+\kv}||f_{ K+\kv-\qv}|}-\sin 4\theta_q \frac{\mathcal{F}^x_{ K+\kv}\mathcal{F}^y_{ K+\kv-\qv}+\mathcal{F}^y_{ K+\kv}\mathcal{F}^x_{ K+\kv-\qv}}{|f_{ K+\kv}||f_{ K+\kv-\qv}|})\non\\
=&-i\sgn \Omega\, 9\lambda_{E_2}^2q^2\pi (1-\cos 6\theta_q)\int \frac{\diff^2 k}{\mathcal{A}_{\text{BZ}}}\delta(\Omega+|f_{K+\kv-\qv}|-|f_{ K+\kv}|)\big(\tanh{\frac{\beta|f_{ K+\kv}|}{2}}-\tanh{\frac{\beta|f_{ K+\kv-\qv}|}{2}}\big)\non\\
=& - i\frac{18\pi\lambda_{E_2}^2 q|\Omega|}{v_F^3\mathcal{A}_{\text{BZ}}}\beta(1-\cos 6\theta_q) \int_{\epsilon>v_Fq/2} \diff \epsilon \frac{\epsilon}{\sqrt{4-(v_F q/\epsilon)^2}}\frac{1}{\cosh^2\beta\epsilon/2}\non\\
=& - i\frac{18\pi\lambda_{E_2}^2 q|\Omega|}{v_F^3\mathcal{A}_{\text{BZ}}}T (1-\cos 6\theta_q)\int_{v_Fq\beta/2}^{\infty} \diff x \frac{x^2}{\sqrt{4x^2-(v_F q \beta)^2}}\frac{1}{\cosh^2x/2}\non\\
\approx& - i\frac{18\pi\lambda_{E_2}^2 q|\Omega|}{2v_F^3\mathcal{A}_{\text{BZ}}}T (1-\cos 6\theta_q) \int_0^{\infty} \diff x \frac{x}{\cosh^2x/2}
=-i\frac{36\pi\lambda_{E_2}^2 q |\Omega|}{v_F^3\mathcal{A}_{\text{BZ}}}T\ln 2 (1-\cos 6\theta_q).
\label{Appeq:PolarizationL1}
\end{align}
The imaginary part of $\im \Pi_{\mathrm{ph}}^{\parallel\parallel}(\qv,\Omega)$ comes from the pole in Eq.~\eqref{eq:D9} using $\im\frac{1}{\Omega+i\delta\sgn \Omega}=-i\pi \delta(\Omega)\sgn\Omega$.
%-i\frac{18\lambda_{E_2}^2 q^2}{v_F^2\mathcal{A}_{\text{BZ}}}\frac{v_s}{v_F}T\log 2 (1-\cos 6\theta_q)
From the first to the second equation, we use
\begin{align}
&\frac{\mathcal{F}^x_{ K+\kv}\mathcal{F}^y_{ K+\kv-\qv}+\mathcal{F}^y_{ K+\kv}\mathcal{F}^x_{ K+\kv-\qv}}{|f_{ K+\kv}||f_{ K+\kv-\qv}|}=-\sin2\theta_q,\non\\
&\frac{\mathcal{F}^x_{ K+\kv}\mathcal{F}^x_{K+\kv-\qv}-\mathcal{F}^y_{ K+\kv}\mathcal{F}^y_{ K+\kv-\qv}}{|f_{ K+\kv}||f_{ K+\kv-\qv}|}=\cos 2\theta_q.
\end{align} 
From the penultimate to the last line, we require $v_F q \beta \ll 1$ ($v_F q \ll T$), which optimizes the decay, because otherwise, the integrand is exponentially small as $\frac{1}{\cosh^2 x/2}\ll 1$.

Similarly, we find 
\begin{align}
i\im \Pi^{\perp\perp}(\qv,\Omega)=-i\frac{36\pi\lambda_{E_2}^2 q |\Omega|}{v_F^3\mathcal{A}_{\text{BZ}}}T\ln 2 (1+\cos 6\theta_q).
\label{Appeq:PolarizationT}
\end{align}

%%%%%%%%%%%%%%%%%%%%%%%%%%%%%%%%%%%%%%%%%%%%%%%%%%%%%%%%%

%%%%%%%%%%%%%%%%%%%%%%%%%%%%%%%%%%%%%%%%%%%%%%%%%%%%%%%%%

\bibliography{SpinPhonon_refs}
\end{document}